\pdfoutput=1
\documentclass[11pt,twoside,a4paper,cmspaper,final,collab]{cms-tdr}
\def\svnVersion{c4af0a3}\def\svnDate{2024/07/14}\def\cmsCernNoTag{CERN-EP-2023-113}\def\cmsCernDate{\today}\def\cmsMessage{Published in the Journal of High Energy Physics as \href{http://dx.doi.org/10.1007/JHEP06(2024)187}{\doi{10.1007/JHEP06(2024)187}.}}
\begin{document}\cmsNoteHeader{TOP-21-007}

\renewcommand{\cmsCollabName}{The CMS and TOTEM Collaborations}
\renewcommand{\cmsTag}{CMS-\cmsNUMBER\\&TOTEM 2023-002\\}
\renewcommand{\cmsPubBlock}{\begin{tabular}[t]{@{}r@{}l}&CMS \cmsSTYLE~\cmsNUMBER\\&TOTEM 2023-002\\\end{tabular}}
\renewcommand{\cmsCopyright}{\copyright\,\the\year\ CERN for the benefit of the CMS and TOTEM Collaborations.}
\renewcommand{\appMsg}{See Appendices~\ref{app:collab} and~\ref{app:totem} for the lists of collaboration members}
\renewcommand{\cmslogo}{\includegraphics[height=2cm]{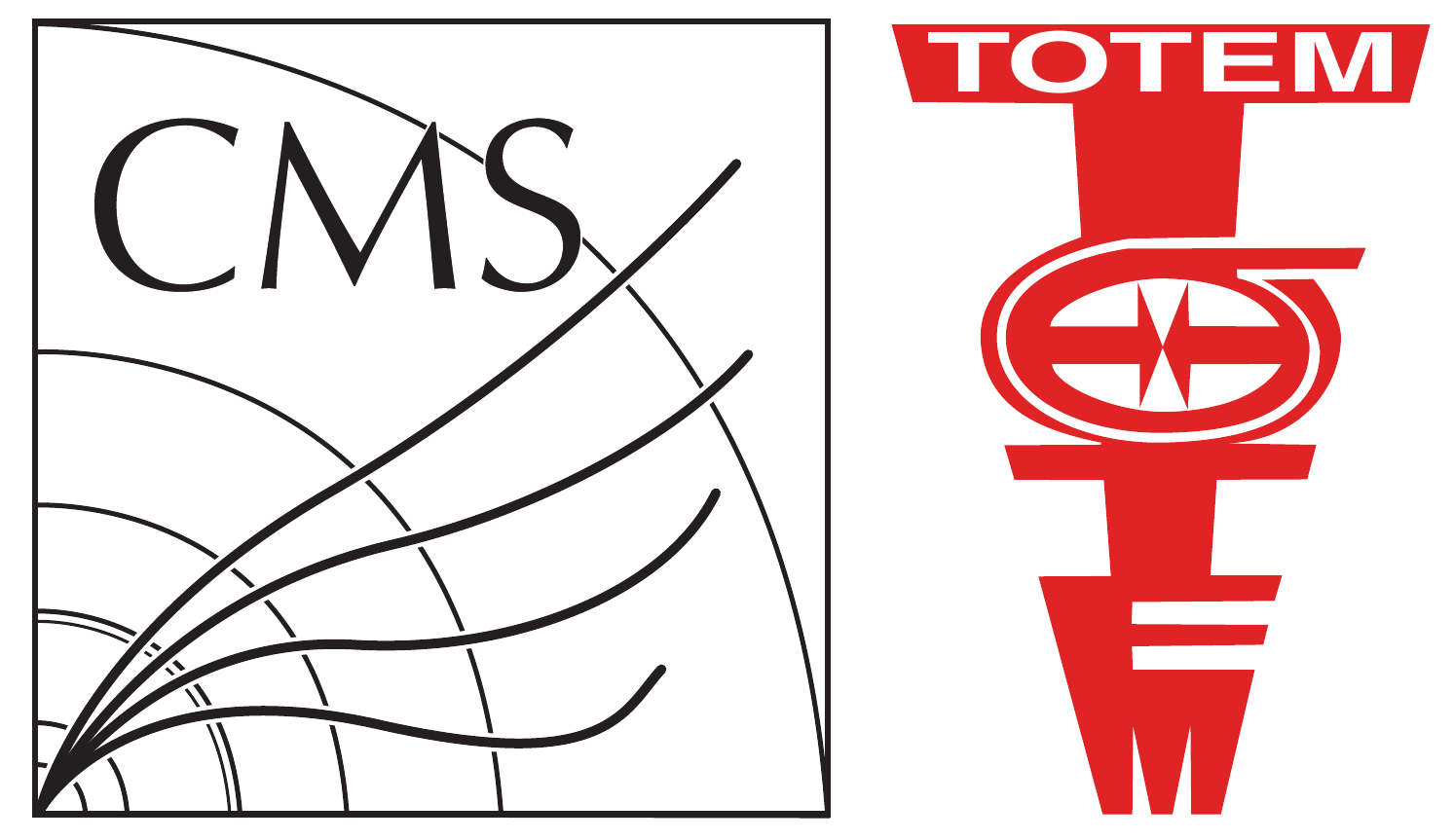}}
\newcommand{\GG}{\ensuremath{\PGg\PGg}\xspace}
\newcommand{\pp}{\ensuremath{\Pp\Pp}\xspace}
\newcommand{\GGtottbar}{\ensuremath{\GG\to\ttbar}\xspace}
\newcommand{\pptopttbarp}{\ensuremath{\pp\to\Pp\ttbar\Pp}\xspace}
\newcommand{\tW}{\ensuremath{\PQt\PW}\xspace}
\newcommand{\leptonjets}{\ensuremath{\text{lepton}+\text{jets}}\xspace}
\newcommand{\ljets}{\ensuremath{\Pell+\text{jets}}\xspace}
\newcommand{\alphaX}{\ensuremath{\alpha_{\mathrm{X}}}\xspace}
\newcommand{\mll}{\ensuremath{m_{\Pell\Pell}}\xspace}
\newcommand{\murad}{\ensuremath{\,\mu\text{rad}}\xspace}
\newcommand{\vecpi}{\ensuremath{\vec{p}_{\mathrm{i}}}\xspace}
\newcommand{\vecpo}{\ensuremath{\vec{p}_{\mathrm{o}}}\xspace}
\newcommand{\abseta}{\ensuremath{\abs{\eta}}\xspace}
\newcommand{\ee}{\ensuremath{\Pe\Pe}\xspace}
\newcommand{\emu}{\ensuremath{\Pe\PGm}\xspace}
\newcommand{\mumu}{\ensuremath{\PGm\PGm}\xspace}
\newcommand{\mW}{\ensuremath{m_{\PW}}\xspace}
\newcommand{\mtop}{\ensuremath{m_{\PQt}}\xspace}
\newcommand{\mtt}{\ensuremath{m_{\ttbar}}\xspace}
\newcommand{\mtopref}{\ensuremath{m^{\text{ref}}_{\PQt}}\xspace}
\newcommand{\mrecotop}{\ensuremath{m^{\text{reco}}_{\PQt}}\xspace}
\newcommand{\mrecotbar}{\ensuremath{m^{\text{reco}}_{\PAQt}}\xspace}
\newcommand{\nvtx}{\ensuremath{n_{\text{vtx}}}\xspace}
\newcommand{\Pmc}{\ensuremath{P^{\text{MC}}}\xspace}
\newcommand{\Pdata}{\ensuremath{P^{\text{data}}}\xspace}

\DeclareRobustCommand{\PVpr}{{\HepParticle{\PV}{}{\prime}}\Xspace}

\cmsNoteHeader{TOP-21-007}

\title{Search for central exclusive production of top quark pairs in proton-proton collisions at \texorpdfstring{$\sqrt{s} = 13\TeV$}{sqrt(s) = 13 TeV} with tagged protons}

\author*[cern]{The CMS and TOTEM Collaborations}

\date{\today}

\abstract{A search for the central exclusive production of top quark-antiquark pairs (\ttbar) is performed for the first time using proton-tagged events in proton-proton collisions at the LHC at a centre-of-mass energy of 13\TeV. The data correspond to an integrated luminosity of 29.4\fbinv. The \ttbar decay products are reconstructed using the central CMS detector, while forward protons are measured in the CMS-TOTEM precision proton spectrometer. An observed (expected) upper bound on the production cross section of 0.59 (1.14)\unit{pb} is set at 95\% confidence level, for collisions of protons with fractional momentum losses between 2 and 20\%.}

\hypersetup{%
pdfauthor={CMS Collaboration},%
pdftitle={Search for central exclusive production of top quark pairs in proton-proton collisions at sqrt(s) = 13 TeV with tagged protons},%
pdfsubject={CMS},%
pdfkeywords={CMS, top quark, exclusive production, CT-PPS}}

\maketitle

\section{Introduction}

Top quarks are copiously produced in proton-proton (\pp) collisions at the CERN LHC.
At LHC energies, the dominant production mode is via strong
interaction processes, resulting in the production of top quark-antiquark pairs (\ttbar).
The LHC experiments have measured the inclusive \ttbar production cross
section at various centre-of-mass
energies, using different top quark decay channels~\cite{CMS:2013yjt, ATLAS:2014nxi, CMS:2015auz, CMS:2016csa, CMS:2016yys,
 ATLAS:2017jkf, ATLAS:2017wvi, LHCb:2018usb, CMS:2018fks, CMS:2019snc,
 ATLAS:2020ccu, ATLAS:2020aln, CMS:2021vhb, CMS:2021gwv, ATLAS:2022aof,
 ATLAS:2022jbj, ATLAS:2022opp, ATLAS:2023gsl, CMS:2023qyl, ATLAS:2023slx}.
Top quarks can also be produced singly in electroweak processes in
three different modes known as $t$ channel,
$s$ channel, and \PW-associated production (\tW). The ATLAS and CMS
Collaborations have observed or reported evidence for single top quark
production in all three modes at several centre-of-mass energies
\cite{Aaboud:2019pkc, Aaboud:2016ymp, Sirunyan:2019hqb, Aaboud:2016lpj}.

\begin{figure}[!htp]
  \centering
  \includegraphics[width=0.4\textwidth]{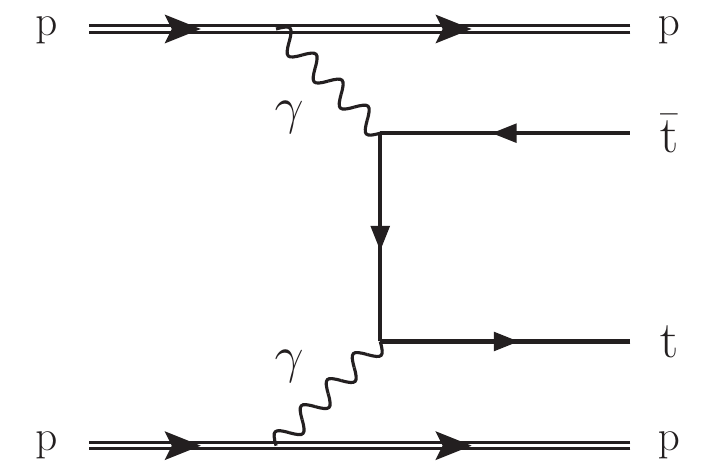}
  \caption{Leading Feynman diagram for \ttbar central exclusive production via
  \GG fusion.}
  \label{fig:feynman}
\end{figure}

A different mechanism can lead to the production of \ttbar pairs in
\pp scattering via the exchange of colourless particles, such as photons (\PGg)
or pomerons. In this case, one or both protons may remain intact after
the interaction, while part of their energy is used to produce the \ttbar
pair. The process where the two protons survive the collision, \pptopttbarp, is called
central exclusive production. It receives contributions from quantum
electrodynamics (QED) and quantum chromodynamics (QCD) diagrams~\cite{Khoze:2001xm}.
The diagram with \GG fusion, sketched
in Fig.~\ref{fig:feynman}, is expected to dominate in the phase space region accessible
to forward proton detectors at the LHC~\cite{Khoze:2001xm, Fichet:2014uka}; the pomeron-pomeron
fusion, which can be described at the lowest order in perturbation theory as a
colour-singlet two-gluon exchange, as well as the photoproduction (\PGg{}-pomeron)
process, give negligible contributions in comparison.

Predictions for central \ttbar exclusive production in the framework of
the standard model (SM) are available, including both QED and QCD
contributions~\cite{Fichet:2014uka, deFavereaudeJeneret:2009db, dEnterria:2009cwl,
Fayazbakhsh:2015xba, Luszczak:2018dfi, Goncalves:2020saa,
Shao:2022cly}. A critical element, in particular in the case of
strong interaction processes, is the evaluation of the so-called proton survival
probability. This is the probability that no additional soft
interactions between the spectator partons of the colliding protons take place,
which can lead to energy loss and/or break up of the interacting protons.
For the \GG fusion process this value is close to unity,
while it is limited to a few percent for the QCD processes. The cross section for the
$\pp \to \Pp\GG\Pp \to
\Pp\ttbar\Pp$ process (referred to as \GGtottbar) amounts to $0.22\pm 0.05\unit{fb}$ including
next-to-leading-order (NLO) perturbative QCD corrections~\cite{Shao:2022cly}.
While the observation of the central exclusive
production of \ttbar pairs is only expected to become possible at the
high-luminosity LHC~\cite{CMS:2021ncv}, contributions from physics beyond the
SM could enhance the production cross section, making it
detectable with the data collected so far. In particular, this production mechanism is
sensitive to the $\PQt\PGg$ vertex, which makes it suitable for interpretations
in the context of Effective Field Theory~\cite{Aguilar-Saavedra:2018ksv} or
anomalous couplings~\cite{Fayazbakhsh:2015xba, Baldenegro:2022kaa}.
This offers complementary information to processes like $\ttbar\PGg$ production,
measured by CMS and ATLAS at 13\TeV~\cite{CMS:2022lmh, CMS:2021klw, ATLAS:2018sos,
ATLAS:2020yrp}. This process is also sensitive to models that incorporate
extra spatial dimensions~\cite{Inan:2011zz}.

This paper reports on a search for central exclusive \ttbar production at the
LHC, carried out by reconstructing the top quarks from their
decay products in the CMS central detector, and looking for the
presence of two forward protons with the CMS-TOTEM precision proton
spectrometer (CT-PPS)~\cite{Albrow:2014lrm}. Each top quark decays almost always
to a \PW boson and a bottom quark. At least one of the two \PW bosons from top
quark decays is reconstructed in the leptonic ($\Pe\PGne$ or $\PGm\PGnGm$) channel
(including $\PW\to\PGt\PGnGt$ decays where the tau lepton decays leptonically), while the
other \PW boson is reconstructed either in the leptonic or hadronic decay
mode. Throughout the paper,
the events where both top quarks decay in the leptonic
channel are referred to as dileptonic, while events with one top quark decaying
leptonically and the other hadronically are referred to as \leptonjets (\ljets).
The two scattered protons are detected by CT-PPS, one on each
side of the interaction region. The analysis is based on data collected in 2017.

The paper contains seven sections. Section~\ref{sec:exp} briefly illustrates the
CMS detector, the CT-PPS experimental setup, and the reconstruction of basic objects.
Section~\ref{sec:samples} specifies the data and simulation
samples used in the analysis. Section~\ref{sec:analysis} outlines the analysis
strategy, and details its various steps. Section~\ref{sec:syst} is
devoted to the treatment of systematic uncertainties. Section~\ref{sec:results}
describes the statistical analysis and presents the results.
The paper is closed with a summary in Section~\ref{sec:summary}.

\section{Experimental setup and particle reconstruction}
\label{sec:exp}

\subsection{The CMS detector and the CMS-TOTEM precision proton spectrometer}
\label{subsec:cms-pps}

The central feature of the CMS apparatus is a superconducting solenoid
of 6\unit{m} internal diameter, providing a magnetic field of
3.8\unit{T}. Within the solenoid volume are a silicon pixel and strip
tracker, a lead tungstate crystal electromagnetic calorimeter (ECAL),
and a brass and scintillator hadron calorimeter, each composed
of a barrel and two endcap sections. Forward calorimeters extend the
pseudorapidity ($\eta$) coverage provided by the barrel and endcap
detectors. Muons are detected in gas-ionisation chambers embedded in
the steel flux-return yoke outside the solenoid.
A more detailed description of the CMS central detector, together with a
definition of the coordinate system used and the relevant kinematic
variables, can be found in Ref.~\cite{Chatrchyan:2008aa}.

\begin{figure}[!b]
  \centering
  \includegraphics[width=\textwidth]{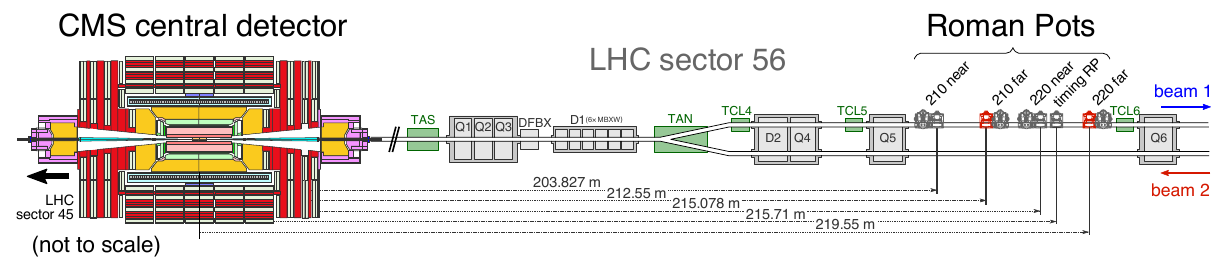}
  \caption{A schematic layout of one arm of CT-PPS along the LHC beam
  line. The RPs shown in red host the detectors used in this analysis.}
  \label{fig:ppsLayout}
\end{figure}

The CT-PPS detector is an array of movable, near-beam devices,
called Roman Pots (RPs), enclosing tracking or timing detectors,
and installed along the LHC beam line at about 210\unit{m} from the
CMS interaction point (IP), on both sides, in LHC sectors 45 (``arm~0'') and 56 (``arm~1''). A sketch of the
system layout for one arm is shown in Fig.~\ref{fig:ppsLayout}. During normal data taking,
detectors are inserted horizontally, their edges approaching the beam
as close as 2--3\mm from its nominal orbit, in
order to reconstruct the flight path of intact scattered protons coming from the
IP. As insufficient information was available from the timing detectors in 2017, only data from the tracking stations are used in this analysis.
In the 2017 configuration, one tracking station per side was
equipped with silicon strip detectors~\cite{Ruggiero:2009zz} and one with silicon pixel
detectors~\cite{Ravera:2016odg}, at a distance of about 213 (``210~far'') and 220\unit{m} (``220~far'') from the IP,
respectively. They can provide up to five and up to six measured points per
track, respectively. Each strip tracker allows the reconstruction of
at most one proton track per event; if hits compatible with more than one track are
reconstructed in at least one strip tracker, the event is discarded, to avoid
ambiguities arising from wrong combinations of orthogonal strips. Each pixel
tracker allows the reconstruction of multiple tracks per event, up to 10.

\subsection{Particle reconstruction}
\label{subsec:particleReco}

In CMS, object reconstruction is based on the
particle-flow algorithm~\cite{CMS-PRF-14-001}, which aims at reconstructing
and identifying each individual particle in an event, with an optimised
combination of information from the various detector elements.

The electron momentum is estimated by combining the energy measurement
in the ECAL, including all bremsstrahlung photons spatially
compatible with originating from the electron track, with the momentum
measurement in the tracker.
The transverse momentum (\pt) resolution ranges from 1.6 to 5\%
for electrons with $\pt \approx 45\GeV$ from $\PZ\to\EE$ decays~\cite{CMS:2020uim}.

The muon momentum is obtained from the curvature of the corresponding
track. Matching muons to tracks measured in the silicon tracker results
in a relative \pt resolution of 1\% in the barrel and 3\% in the endcaps~\cite{CMS:2018rym},
for muons with \pt up to 100\GeV.

The primary vertex (PV) is selected using tracking information only: vertices with at least four tracks
and a longitudinal distance of less than 24\cm from the centre of the detector are selected. From these candidates,
the PV is taken as the one with largest scalar sum of associated particle \pt, as described in Section 9.4.1 of Ref.~\cite{Contardo:2015bmq}.

Jets are clustered from reconstructed particles using the anti-\kt
algorithm~\cite{Cacciari:2008gp, Cacciari:2011ma} with a distance
parameter of 0.4.
The jet momentum is determined as the vectorial sum of all particle
momenta in the jet, and is found from simulation to be, on average,
within 5--10\% of the true momentum over the whole \pt spectrum and
detector acceptance. To mitigate effects from additional \pp interactions within the
same or nearby bunch crossings (``pileup''), tracks identified to be
originating from pileup vertices are discarded, and an offset
correction is applied to correct for remaining contributions~\cite{Cacciari:2008gn,Cacciari:2007fd}. Jet
energy corrections are derived from simulation studies so that the
average measured energy of jets becomes identical to that of particle-level jets. In-situ measurements of the momentum balance in dijet,
\ensuremath{\text{photon}+\text{jet}}\xspace, \ensuremath{\PZ+\text{jet}}\xspace, and multijet events are used to determine any
residual differences between the jet energy scale in data and in
simulation, and appropriate corrections are
applied~\cite{Khachatryan:2016kdb}.
The jet energy resolution amounts typically to 15--20\% at 30\GeV,
10\% at 100\GeV, and 5\% at 1\TeV.

The missing transverse momentum vector \ptvecmiss is computed as the
negative vector sum of the \pt of all the particle-flow candidates
in an event, and its magnitude is denoted as
\ptmiss~\cite{Sirunyan:2019kia}. The vector \ptvecmiss is modified to account
for corrections to the energy scale of the reconstructed jets in the
event.

Intact protons emerging from interaction vertices at small angles are
detected by CT-PPS, either with a single RP station
(pixels or strips), or the combination of the information from two stations in the same
arm (multi-RP reconstruction). The latter features superior
resolution, thanks to the lever arm between the two stations, while it
suffers from lower efficiency because of the
double-track requirement. In this analysis, only multi-RP proton
candidates are used.
The proton reconstruction efficiency is evaluated as the product of three different
contributions~\cite{TOTEM:2022vox}. The first one is the efficiency of the strip detectors, locally
degrading in time because of radiation damage. The second contribution is the multi-RP
reconstruction efficiency, which combines the acceptance of protons propagating
between the near and far stations, the pixel detector efficiency (similarly affected
by radiation damage), and the efficiency of the reconstruction algorithm. Values
for the combination of these two effects are determined, as functions of the position
of the track in the transverse plane, for each of five data-taking periods (``eras'').
Finally, the efficiency of the single-track requirement in the strips mentioned in
Section~\ref{subsec:cms-pps} is taken into account by applying scaling factors,
derived globally per arm, for each era. This is the most
significant contribution to proton reconstruction efficiency, with
values below 50\% for the periods with the highest instantaneous luminosity.

The kinematic state of the proton is characterised by the fractional momentum
loss, defined as $\xi = (\abs{\vecpi} - \abs{\vecpo})/\abs{\vecpi}$, where
\vecpi and
\vecpo are the momenta of the incoming and outgoing protons, respectively.
The value of $\xi$ is derived from the measured slopes and
intercepts of the outgoing proton along with detailed knowledge of the LHC magnetic field.
Dedicated alignment and
calibration procedures are in place for different fills and LHC optics
setup~\cite{TOTEM:2022vox}. The detector acceptance as a function
of $\xi$ is determined by the geometry of the detectors and the LHC
collimators, and also depends on the specific LHC settings:
in 2017, most detectable protons had $\xi$ values in the range
$0.02 < \xi < 0.15$~\cite{TOTEM:2022vox}. Those used in the analysis
are required to be within fiducial regions in the $\xi - \theta^{*}_{x}$ plane
(with $\theta^{*}_{x}$ denoting the proton scattering angle in the horizontal plane at the IP)
where the efficiency can be reliably determined.

\section{Data and simulation samples}
\label{sec:samples}

This analysis uses data collected in 2017 considering only runs where all CT-PPS strip
and pixel detectors were operational, which corresponds to an integrated luminosity of
29.4\fbinv~\cite{CMS:2018elu, CMS:LUM-17-003}.
The beam crossing angle at the
IP, \alphaX, defined here as the angle between the LHC axis and one of the beams,
was set at different values, with most data
being recorded at $\alphaX = 120$,
130, 140, or 150\murad. The remaining data, corresponding to less than 1\fbinv,
are not included in this analysis.

To simulate the signal and background processes, different Monte Carlo (MC)
event generators are used. For all processes, the response of the
central CMS detector is simulated using the \GEANTfour
package~\cite{Agostinelli:2002hh}.

A \GGtottbar signal sample is produced at leading order using
\textsc{fpmc}~\cite{Boonekamp:2011ky} as the matrix element generator,
with the equivalent photon approximation for the photon
flux~\cite{Budnev:1975poe} and a proton survival probability of 0.9~\cite{Baldenegro:2022kaa, CMS:2021ncv}.
Events are generated for $0.02 < \xi < 0.20$. Top quark decays are
simulated with \textsc{MadSpin}~\cite{Artoisenet:2012st},
selecting dilepton and \ljets decays. The outgoing protons are
propagated through the beamline from the IP to the RPs using a fast forward-proton
simulation that includes beam-divergence and vertex smearing at the IP as well as
the beam crossing angle dependence~\cite{TOTEM:2022vox}. Hits in the CT-PPS
detectors are simulated taking into account aperture limitations for a
given crossing angle, and sensor acceptance and resolution.
The efficiency is accounted for at a later stage by assigning appropriate weights
to the events as discussed in Section~\ref{subsec:models}.
The simulated hits are then used to reconstruct proton
tracks by means of the standard CT-PPS reconstruction algorithms.

Backgrounds arise from a variety of hard processes in combination with two uncorrelated
protons from pileup interactions within the CT-PPS acceptance. The dominant
hard-process background is inclusive \ttbar production. A smaller contribution
comes from single top quark production in the \tW channel
and, for the \ljets channel, from QCD multijet events;
additionally, depending on the \ttbar
decay channel, there are small but non-negligible contributions from \ensuremath{\PV+\text{jet}}\xspace, where \PV is either a \PW or a \PZ boson, and Drell--Yan events.
Other possible background sources such as inclusive $\PV\PVpr$ production and
other single-top production channels have been found to have negligible impact
and are not considered further in the analysis.

The inclusive \ttbar sample is simulated at NLO
precision using the
\POWHEG (v2.0)~\cite{Nason:2004rx,Frixione:2007vw,Alioli:2010xd} event
generator. The inclusive \ttbar production
cross section is scaled before the fit to the best available theoretical prediction
at next-to-next-to-leading-order (NNLO) in
QCD, amounting to 832\unit{pb}~\cite{Czakon:2013goa}.
For all background sources containing top quarks, the \pt spectra of
top quarks in simulated samples are reweighted according to
predictions at NNLO QCD accuracy~\cite{Czakon:2017wor}.
For both signal and background event generation, a top quark mass of
172.5\GeV is assumed.

For all processes, the parton showering and hadronisation are
simulated using \PYTHIA8.2~\cite{Sjostrand:2014zea} with
the CP5 underlying event tune~\cite{CMS:2019csb}. The
NNPDF3.1~\cite{Ball:2017nwa} NNLO parton distribution
functions (PDFs) are used.

No simulated sample is used to evaluate the contribution of the
QCD multijet background. Instead, a purely data-driven method is
applied, as described in Section~\ref{subsec:multijet}.

\section{Analysis strategy}
\label{sec:analysis}

The analysis is conducted independently for the events in the dilepton decay channel and
for those in the \ljets decay channel. The resulting distributions from the two channels
are used as input to a common maximum likelihood fit, and a combined result is extracted.

\subsection{Event selection}
\label{subsec:evSel}

Events of interest are selected by CMS using a two-tiered trigger system. The
first level, composed of custom hardware processors, uses
information from the calorimeters and muon detectors to select events
at a rate of around 100\unit{kHz}~\cite{Sirunyan:2020zal}. The second level, known as the
high-level trigger (HLT), consists of a farm of processors running a
version of the full event reconstruction software optimised for fast
processing, and reduces the event rate to around 1\unit{kHz} before
data storage~\cite{Khachatryan:2016bia}.

In the dilepton analysis, events are selected using a combination of single-lepton
and dilepton triggers that identify leptons within $\abseta<2.5$. The single-lepton HLT
selection requires the presence of an isolated electron~\cite{CMS:2020uim} (muon~\cite{CMS:2018rym}) reconstructed with
$\pt>35$ (27)\GeV. Alternatively, a 24\GeV requirement is applied for muons within $\abseta<2.1$.
The dilepton HLT selection requires the presence of two isolated
electrons with $\pt>23$ and 12\GeV, two isolated muons with $\pt>17$ and 8\GeV, one
isolated electron with $\pt>23\GeV$ and one isolated muon with $\pt>8\GeV$, or one
isolated muon with $\pt>23\GeV$ and one isolated electron with $\pt>12\GeV$.

In the \ljets analysis, events are selected using a combination of single-lepton
and jet triggers. The single-lepton HLT selection requires the presence of a single isolated electron (muon) with $\pt>35$ (27)\GeV,
reconstructed within $\abseta<2.5$. The remaining selections require the presence of a single electron with $\pt>28\GeV$
and a sum of the \pt of the jets greater than 150\GeV, or the presence of a single electron with $\pt>30\GeV$ and at least
one jet with $\pt>35\GeV$; in both cases, the electron must be reconstructed within $\abseta<2.1$.

Offline, the reconstructed lepton with highest \pt must have
$\pt > 30\GeV$ and, if it is an electron, it must have $\abseta < 2.1$, while if it is a
muon it must have $\abseta < 2.1 (2.4)$ for the dilepton (\ljets) analysis.
In the dilepton analysis, the lepton with the second highest value of \pt must have
$\pt > 20\GeV$ and $\abseta < 2.4$. Additionally, the charged leptons are required to
satisfy specific quality criteria.
A set of scale factors is applied to simulated events as a function of the
lepton \pt and $\eta$ to account for
differences observed in the lepton trigger, reconstruction, and identification efficiency
between data and simulation~\cite{CMS:2020uim, CMS:2018rym}.

Reconstructed jets are required to have $\pt>30$ (25)\GeV in the
dilepton (\ljets) channel, and $\abseta<2.4$. Moreover, the angular distance $\DR = \sqrt{\smash[b]{(\Delta\eta)^2 + (\Delta\phi)^2}}$ between a jet and a
lepton must be greater than 0.4, where $\phi$ is the azimuthal angle in radians.

Jets originating from the hadronisation of \PQb quarks are identified
with the \textsc{DeepCSV} algorithm~\cite{Sirunyan:2017ezt} as \PQb-tagged jets.
The ``medium'' working point is used, corresponding to a typical
efficiency of about 70\% for correctly identified \PQb quark jets, with a
misidentification probability of 12 (1)\% for \PQc quark (gluon or light quark) jets.
Scale factors are applied to the simulated events as a function of the jet \pt and $\eta$
to account for the differences observed in the \PQb
jet identification efficiency between data and simulation.

The final selection in the dilepton channel requires the presence of at
least two leptons, with the two highest \pt leptons having opposite charge; the
dilepton system they form is required to have an invariant mass $\mll > 20\GeV$.
For the events with two reconstructed leptons of the same flavour, \mll is required to be
outside a 30\GeV window around the \PZ boson mass peak: $(\mll <
76\GeV) \cup (\mll > 106\GeV)$.
Events are categorised according to the final-state charged leptons as
\ee, \emu, or \mumu. Only events with at least two \PQb-tagged jets are retained.
In the \ljets channel, the final selection requires the
presence of exactly one lepton (electron or muon),
at least two jets passing the \PQb tagging selection criteria, and at least
two jets failing the \PQb tagging selection criteria.

Both the dilepton and the \ljets analysis require one multi-RP
proton track to be reconstructed in each arm.

The overall efficiency of the selection, including detector
acceptance, is about 2\% for the dilepton channel and 0.8\% for the
\ljets channel.

\subsection{Top quark pair reconstruction}

Full reconstruction of the \ttbar pair can be used to relate
its kinematics to that of the forward protons. In central exclusive
production, the momentum transfer at the interaction vertex is typically
quite small, implying very small values (below 1\GeV) for the transverse momentum of the outgoing protons and, consequently, of the central system.
Moreover, the invariant mass and the rapidity of the central system $X$ are related
to the momentum loss of the protons by the expressions:
\begin{linenomath}
  \begin{align}
    m_X &= \sqrt{s\xi_1\xi_2}, \label{eq:MX}\\
    y_X &= \frac{1}{2}\ln\frac{\xi_1}{\xi_2}, \label{eq:yX}
  \end{align}
\end{linenomath}
where $\sqrt{s}$ is the centre-of-mass energy and $\xi_1$, $\xi_2$ are
the fractional momentum losses of the outgoing protons in the positive
and negative $z$ direction, respectively. The reconstruction of
a \ttbar candidate through its decay chain is
carried out independently for the dilepton and \ljets
channels, in order to take advantage of their different kinematic
properties. In the dilepton channel, the \ttbar system is
reconstructed by means of an analytic method, briefly outlined in the following,
and the resulting \ttbar observables are used as input to the multivariate discriminant
described in Section~\ref{subsec:mva}, together with the kinematic
observables of the tagged protons. In the \ljets channel,
the kinematics matching between the \ttbar system and the tagged
protons is explicitly used as a constraint in a global kinematic fit.

In the dilepton analysis, the two charged leptons and the two
\PQb-tagged jets with the highest \pt are selected.
The association of the leptons with the jets relies
on a kinematic reconstruction algorithm~\cite{Sirunyan:2018ucr} that
also estimates the kinematics of the top quark and antiquark.
The missing transverse momentum is assumed to originate
solely from the two neutrinos in the decay, and the \PW boson and top
quark masses, \mW and \mtop, are constrained to their known
values~\cite{ParticleDataGroup:2022pth}. For both lepton-jet combinations, multiple replicas of the
energy-momentum conservation equations are generated, with
particle momenta varied according to their resolution and the width
of the \PW boson. For each of them, the solution with the
smallest value of the \ttbar invariant mass (\mtt) is chosen, and a weight is
assigned based on the resulting invariant mass of the lepton and \PQb quark jet
system, with the generator-level spectrum as reference.
The weights are then used to obtain weighted
averages of the kinematic observables of the top quark and antiquark.
The combination of leptons and jets
that yields the highest sum of weights is chosen.
This algorithm finds a physical solution in about 90\% of the events
passing the previous selection, both for data and for simulation. For simulated \ttbar events, the correct
association of lepton and \PQb jet is achieved in 70\% of the cases. The events for which no physical solution is found
are not removed, but a fixed, unphysical value is assigned to their \ttbar observables.

In the \ljets analysis, only the \PQb-tagged jets and the
non-\PQb-tagged (denoted `light-flavour') jets
with the highest \pt values are considered:
up to four of each type are selected. Top quark candidates with the \PW
boson decaying leptonically are reconstructed from combinations of a
\PQb-tagged jet, the selected lepton, and a neutrino candidate. The
neutrino candidate is initially reconstructed from the missing transverse momentum, with the longitudinal component
assigned by imposing the constraint $m_{\Pell\PGn} = \mW$. In cases where the two solutions of the resulting quadratic equation are real, the one closest to the longitudinal momentum of the lepton is chosen. Top quark
candidates with the \PW boson decaying hadronically are reconstructed
from combinations of a \PQb-tagged and two light-flavour jets.
The choice of the two \PQb quark jets to be used for top quark and antiquark reconstruction,
and of their association with the other objects, is based on the
invariant mass of the reconstructed \PQt and \PAQt candidates,
\mrecotop and \mrecotbar. The combination
that yields the lowest value of
$\abs{\mrecotop - \mtopref} + \abs{\mrecotbar - \mtopref}$
is selected, where \mtopref is chosen to be
173.1\GeV, from direct measurements
\cite{ParticleDataGroup:2022pth}.
Using this procedure, \PQb quark jets are found to be correctly assigned in 75\% of all
cases.
The kinematic observables of all reconstructed objects are further
corrected by means of a kinematic fit. The momentum components of
the lepton, the four jets, and the neutrino, as well as the fractional
momentum loss of the forward protons, are used as inputs to the fit and allowed to float,
constrained by Gaussian probability distribution functions centred on
their measured values and with the widths equal to the measurement
uncertainties. The longitudinal component of the neutrino momentum is
left free to float in the fit. The \PW
boson mass (\mW) and \mtop are constrained to their known values, and the total
\pt of the \ttbar system is set to zero. Finally, \mtt and the fractional
momentum loss of the protons are required to satisfy Eq.~\eqref{eq:MX},
where $X$ is the \ttbar pair.

\begin{figure}[!ht]
  \centering
  \includegraphics[width=0.46\textwidth]{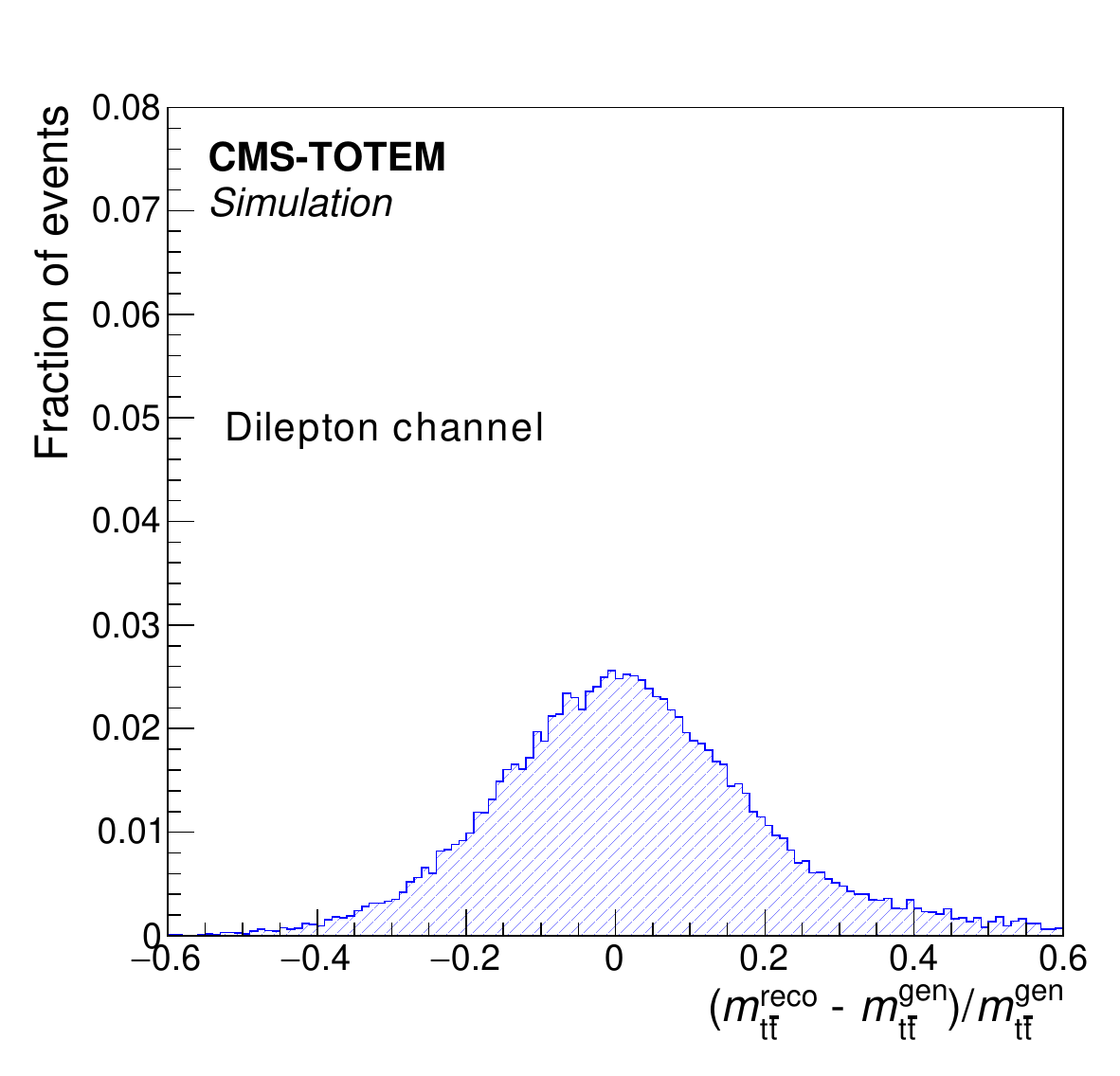}
  \includegraphics[width=0.46\textwidth]{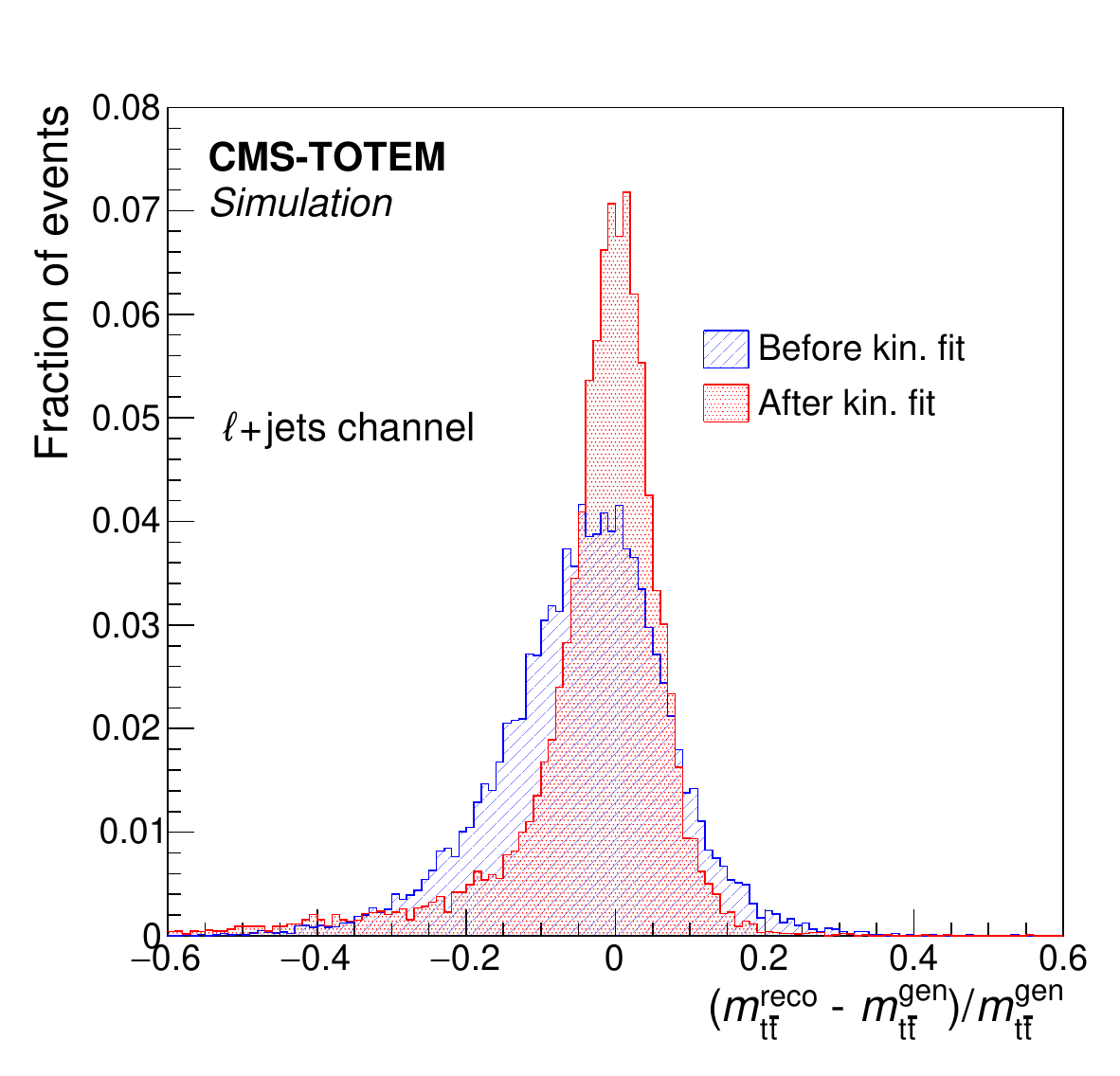}
  \caption{Normalised distribution of the relative
    resolution of the reconstructed \mtt in simulated signal
    events, for the dilepton (left) and \ljets (right)
    analyses. The resolution is shown only for events where the reconstruction is successful.
    For the \ljets decay mode, the hatched blue and the
    dotted red histograms represent the distribution before and after
    applying the kinematic fit, respectively.}
  \label{fig:mttbarRes}
\end{figure}

Figure~\ref{fig:mttbarRes} shows the \mtt resolution achieved for the dilepton and \ljets channels. The poorer resolution
obtained for the dilepton mode, for which the width of the Gaussian core
of the distribution is $\simeq$15\%, is understood from the presence of two
neutrinos in the final state. For the \ljets case, the resolution is
shown before ($\simeq$7.5\%) and after ($\simeq$5\%)
applying the kinematic fitter.

\subsection{Background from multijet events}
\label{subsec:multijet}

For the \ljets analysis, the background originating from
QCD multijet events has been evaluated with a data-driven
approach. The method is based on the observation that the leptons
selected in such events are generally not produced promptly in
the primary interaction, but are rather real leptons from
semileptonic decays of hadrons, or other objects incorrectly
identified as leptons.

Samples of events enriched with nonprompt leptons are created by
imposing looser selection criteria on the lepton. A ``tight-to-loose''
ratio is defined as the ratio of the number of nonprompt lepton events
satisfying the tight (nominal) selection to the number of those
passing the loose ID selections described in Refs.~\cite{CMS:2020uim} and
\cite{CMS:2018rym} but not the tight one. It is evaluated in a data sample mostly
populated by multijet events (``control region'', or CR) and then used to
estimate the number of nonprompt lepton events passing the nominal
\ljets selection (``signal region'', or SR) described in
Section~\ref{subsec:evSel}.

The CR is defined by the same selection criteria as the SR, except for
the requirement that no jet pass the \PQb tagging selection, and that
$\ptmiss < 20\GeV$. Contributions from background sources other than
multijet events are subtracted using the simulated samples. Values of
the tight-to-loose ratio are calculated as a function of the lepton
\pt separately for the two lepton flavours, and then applied to data
in the SR, after all simulated contributions from prompt-lepton
background sources have been subtracted.

This method can be used to obtain the distribution of any kinematic
variable for the nonprompt lepton component, as well as of the
multivariate discriminant used for signal extraction described in Section~\ref{subsec:mva}. For the latter, the resulting
shape is observed to be consistent, within statistical
uncertainties, with that from the dominant inclusive \ttbar
background. Since the inclusive \ttbar normalisation is a free parameter
in the final fit described in Section~\ref{sec:results}, separately for
the dilepton and the \ljets channels, and the contribution
of the nonprompt lepton component is estimated to be much smaller
(about 13\%), an independent QCD multijet background contribution is not included in the final fit.

\subsection{Signal and background models}
\label{subsec:models}

The presence of multiple proton interactions within the same LHC bunch
crossing results in the superposition of objects from different
PVs both in the central CMS apparatus and in CT-PPS.
The probability to have at least one proton in the acceptance of a
given arm of CT-PPS, for any bunch crossing,
ranges from 40 to 70\% depending on the LHC optics
settings and instantaneous luminosity. However, while the pileup activity
in the central detector can be modelled with adequate accuracy, no
simulation has been validated so far for
protons from uncorrelated diffractive events, where the \pp
interaction is mediated by strongly interacting colour-singlet exchange. As a
consequence, in the MC samples, background events contain no forward protons,
while signal events contain exactly two forward protons on opposite sides
(though not necessarily within the acceptance).

The presence of pileup protons, uncorrelated with the event
reconstructed in the central detector, has two effects:
\begin{itemize}
  \item a background event may be selected because exactly one random
    proton per arm has been reconstructed in CT-PPS;
  \item a signal event may be rejected because of the multiple proton
    reconstruction inefficiency, or it may be wrongly reconstructed because a
    background proton is selected instead of the signal one that went
    undetected as a result of detector inefficiency or limited acceptance.
\end{itemize}

In order to correctly take these effects into account, a pool of
forward proton pairs reconstructed in the collision data is collected to be used
as a sample of pileup protons, from events subject to the same requirements of the
nominal selection (including one reconstructed proton in each CT-PPS arm) except for those on \PQb-tagged jets.
In the procedure outlined below, the proton reconstruction efficiency
discussed in Section~\ref{subsec:particleReco} is considered as a function of $\xi$.
Moreover, the probability of having zero (not including multitrack inefficiency) or one proton
reconstructed in each arm is taken from the same studies~\cite{TOTEM:2022vox}.
Because the detector and beam
conditions varied significantly throughout the data taking, both the
forward proton pools and the efficiency/probability values are considered
separately for each of the five eras and, except
for the reconstruction efficiency, for four
values of the beam crossing angle \alphaX at the IP (120, 130, 140,
and 150\murad).

For each simulated event, a pair of protons is selected from the pool
according to the relative normalisation of the (era, \alphaX)
samples. Then, the following procedure is applied:
\begin{itemize}
  \item for background events, the proton pair is added and a weight
    corresponding to the probability of reconstructing one
    proton in each arm is assigned;
  \item for signal events, the number of reconstructed protons is
    first determined according to the detector acceptance and a random
    correction based on the multi-RP reconstruction efficiency. If only one of
    the original protons is left, the other is replaced with one from
    the pool, and an appropriate weight is assigned to the event,
    according to the probability of ending up with exactly one proton in that arm.
    Events in which neither forward proton is reconstructed
    are treated in the same way as background events, as described above.
\end{itemize}

In order to match the pileup conditions for simulated events to those in the
collision data, a further reweighting procedure is applied to
simulated events, based on the number of reconstructed interaction
vertices. The normalised distribution of this number for a given simulated sample,
$\Pmc(\nvtx)$, and that for the data in each of the
20 $(\text{era},\alphaX)$ regions,
$\Pdata(\nvtx \mid \text{era},\alphaX)$, are determined. A
weight $w_\text{PU} = \Pdata(\nvtx \mid \text{era},\alphaX)/\Pmc(\nvtx)$
is assigned depending on the sampled region.

\begin{figure}[!ht]
  \centering
  \includegraphics[width=0.46\textwidth]{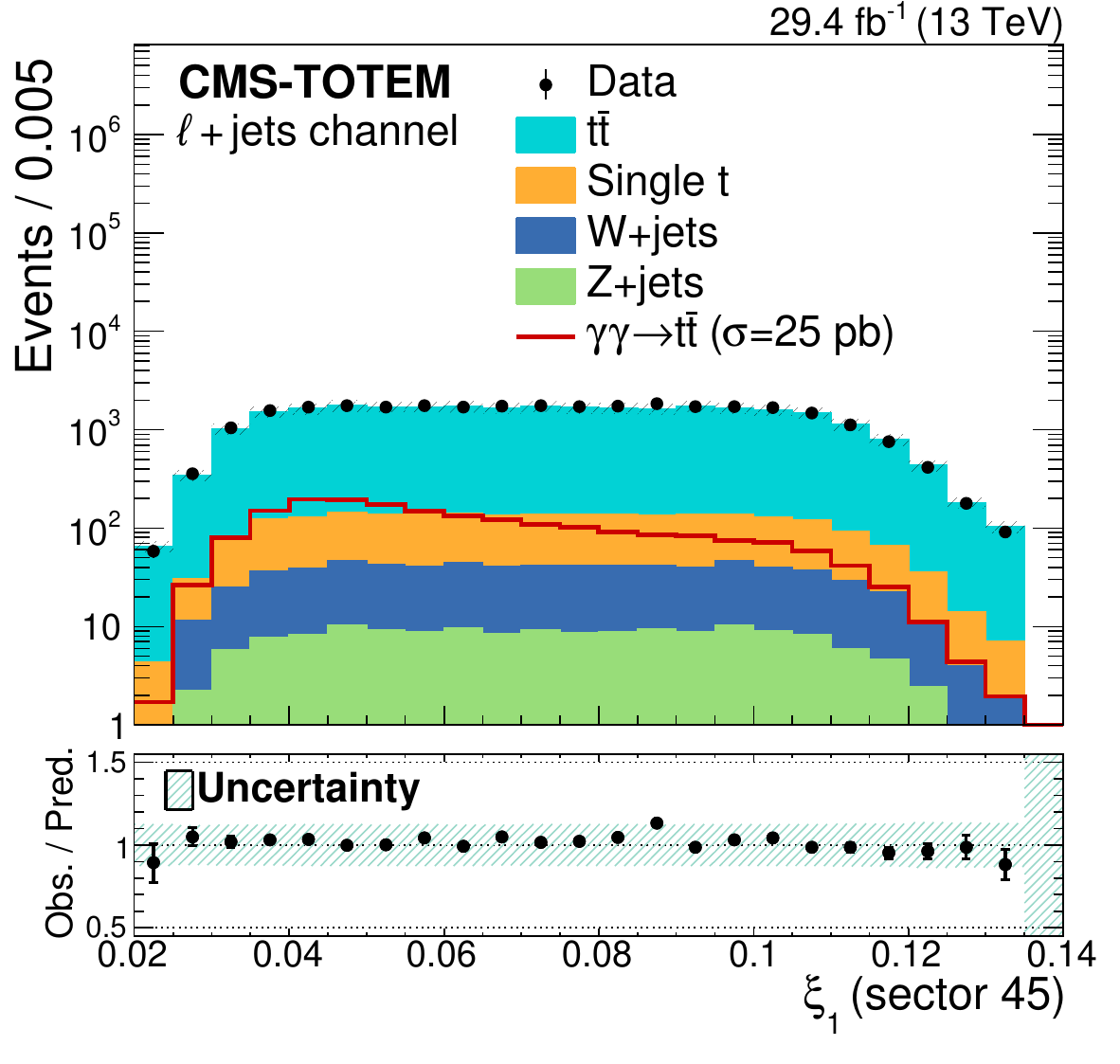}
  \includegraphics[width=0.46\textwidth]{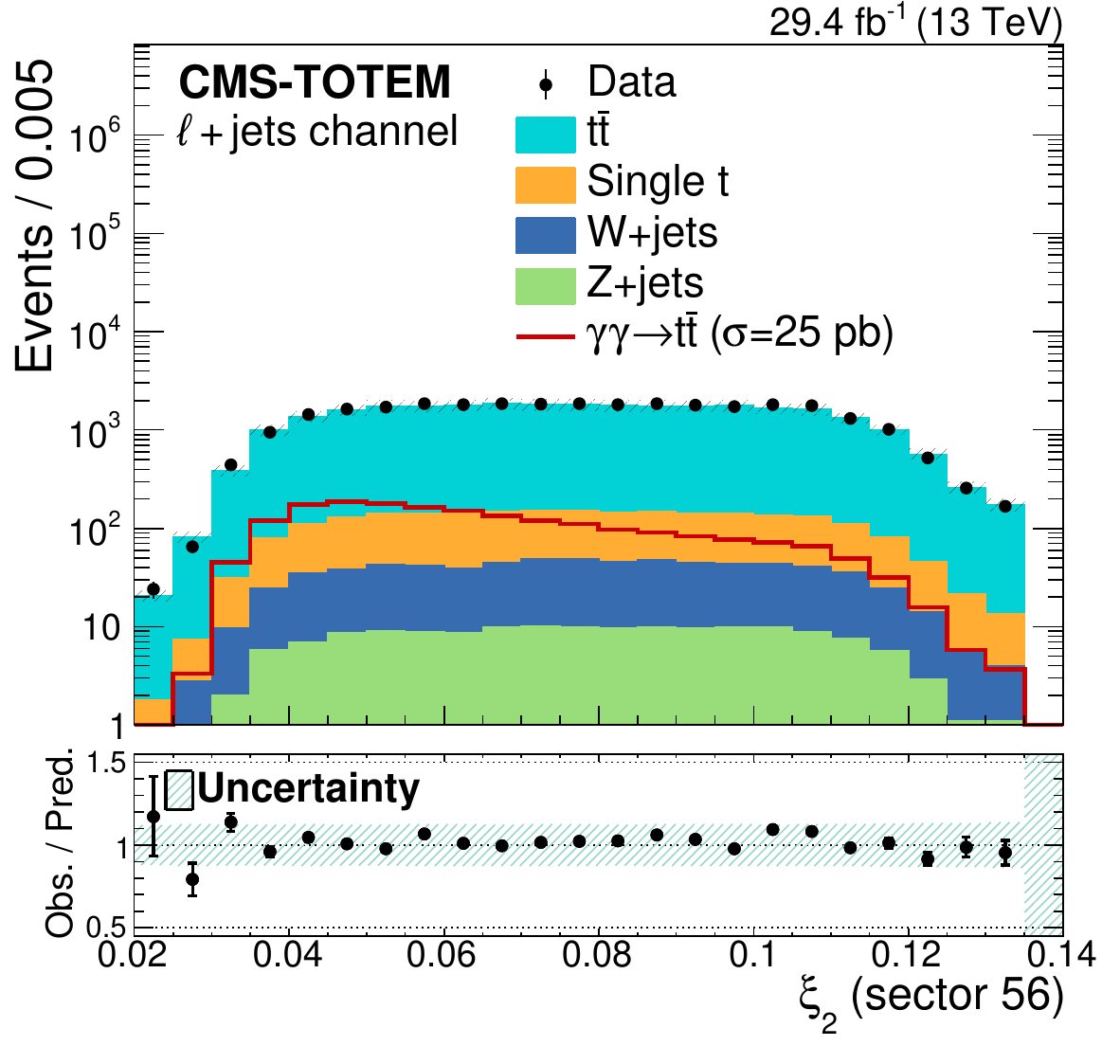}
  \caption{Distribution of $\xi$
    in data and background simulated samples after pileup proton
    mixing and pileup reweighting, in the \ljets channel.
    Protons in CT-PPS arm~0 (left) and arm~1 (right), as
    defined in the text.
    The solid histograms show the expected background contributions,
    while the red open histograms show the expected signal shapes, normalised
    to a cross section of 25\unit{pb}, approximately $10^5$ larger than the
    SM cross section prediction from Ref.~\cite{Shao:2022cly}; points
    with statistical error bars represent collision data.
    The lower panels show the data-to-prediction ratios; the hatched
    bands represent the relative uncertainty in the predictions.}
  \label{fig:mc-data_xi}
\end{figure}

To assess the validity of the background model obtained from this
procedure, the distributions of various event variables in data and
simulated samples are compared, and very good
agreement is observed. Figure~\ref{fig:mc-data_xi} shows the overall distribution of
$\xi$ in each arm of CT-PPS for the \ljets decay mode.

\subsection{Multivariate analysis}
\label{subsec:mva}

In order to enhance the signal content of the selected samples,
information from variables showing discriminating power against
background sources is efficiently exploited by means of multivariate
analysis techniques. For both the dilepton and the \ljets
channels, a boosted decision tree (BDT)
algorithm~\cite{Breiman:1984jka} is used, implemented with the TMVA
toolkit~\cite{TMVA:2007ngy}. The training samples
consist of simulated signal events with both protons reconstructed, and
simulated inclusive \ttbar production events, by far the largest source of
background, with two pileup protons added from collision data, as described in
the previous section. In general, effective discrimination is mostly achieved by
exploiting the absence of extra jets in the exclusive
production event, and the kinematic closure when including both the forward
protons and the centrally produced objects. Because of the different
final products in the two final states and their related kinematics, the specific choice of the
discriminating variables is different for the two decay modes. For each decay mode,
a large set of variables was initially tested, and then reduced to a smaller set
through optimisation, where the most performant and uncorrelated variables were selected.

For the dilepton decay mode, the following 15 kinematic variables are used:
the mass and the rapidity of the central system reconstructed both from
the \ttbar decay products and from proton kinematics
(Eqs.~\eqref{eq:MX} and \eqref{eq:yX}); \ptmiss; the invariant mass
and the angular distance \DR\ of the two leptons; $\abs{\Delta
\phi}$ of the two selected \PQb-tagged jets; the rapidity of the system
formed by the two \PQb quark jets and the two leptons, and the sum of the
absolute values of their individual rapidities; the rapidity of the
system formed by all other reconstructed jets, and the sum of the
absolute values of their individual rapidities; the squared energy sum
for all objects used for the \ttbar reconstruction; the minimum absolute value
of the rapidity difference for any two systems formed by a lepton and
a \PQb-tagged jet; and the number of light-flavour jets.

For the \ljets decay mode, the following 10 kinematic variables are
used: the number of light-flavour jets and of \PQb-tagged jets; the sum of the
invariant mass of all jets; the total energy of all light-flavour jets; the
mean \DR\ for all pairs of light-flavour jets; the total energy of all
extra jets (not used for \ttbar reconstruction); the lepton momentum
and a variable quantifying its isolation from other particles in the event~\cite{CMS:2020uim, CMS:2018rym};
\mtt; the difference in central system rapidity
reconstructed from the \ttbar and the \pp systems
(Eq.~\eqref{eq:yX}); and the $\chi^2$ of the kinematic fit.

\begin{figure}[!ht]
  \centering

  \includegraphics[width=0.32\textwidth]{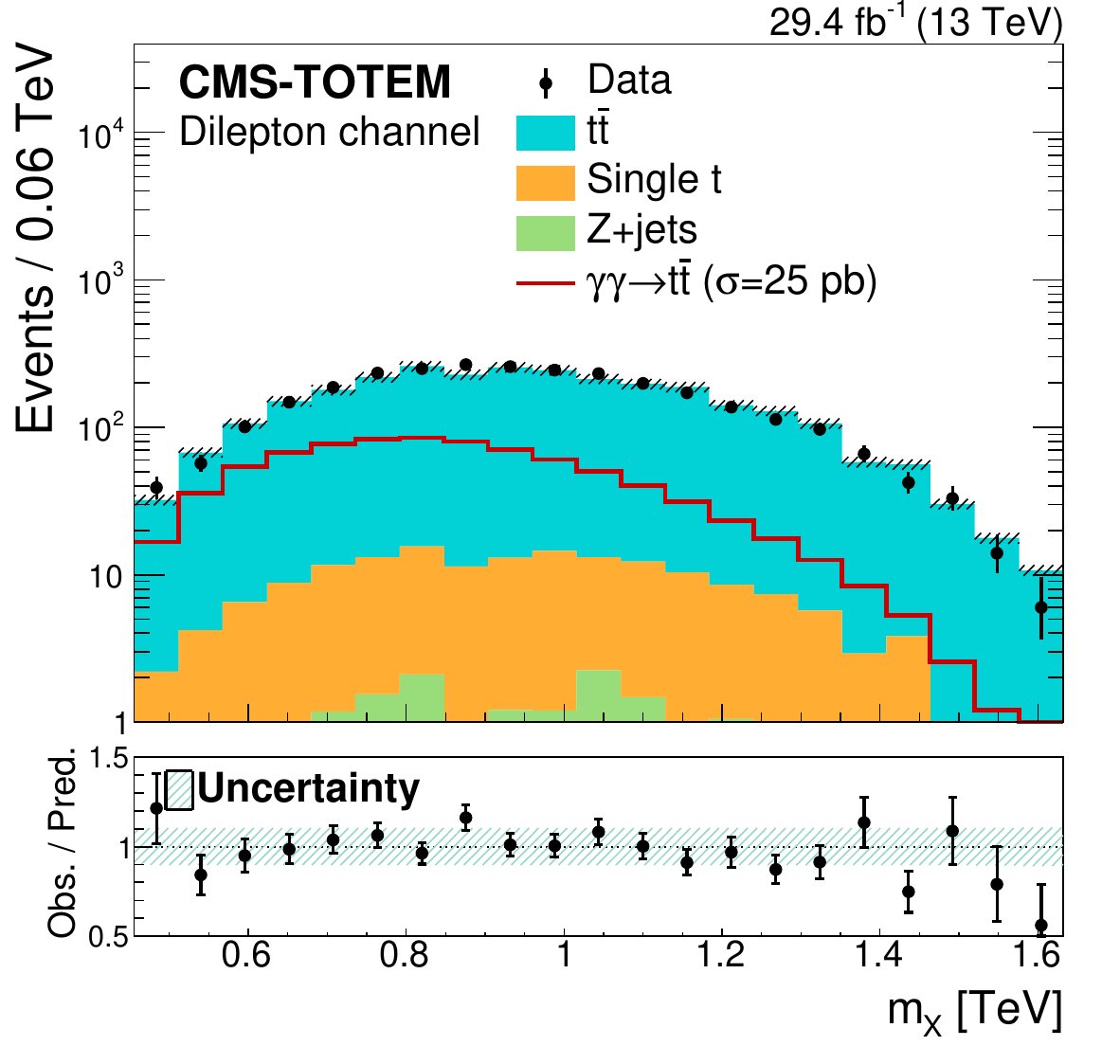}
  \includegraphics[width=0.32\textwidth]{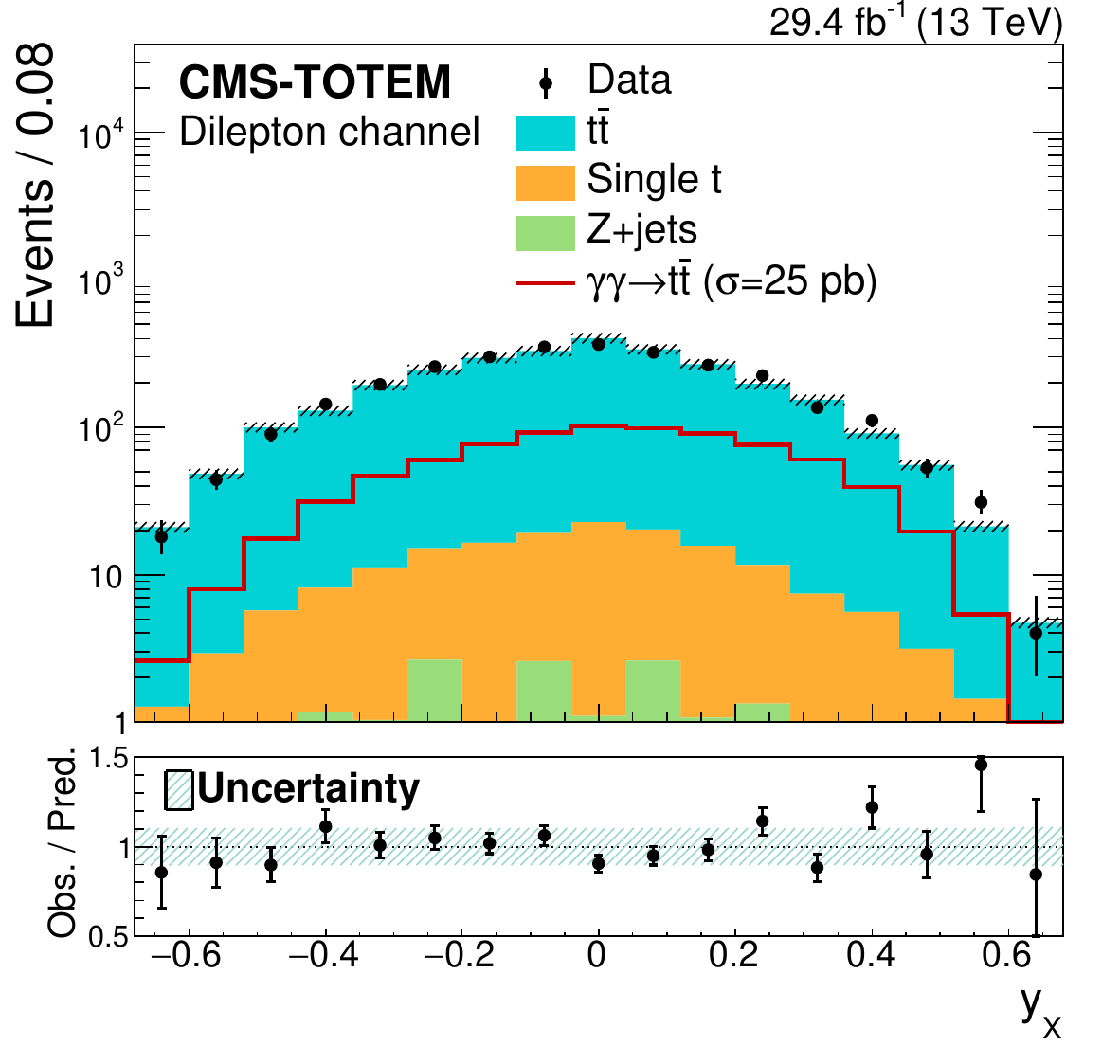}
  \includegraphics[width=0.32\textwidth]{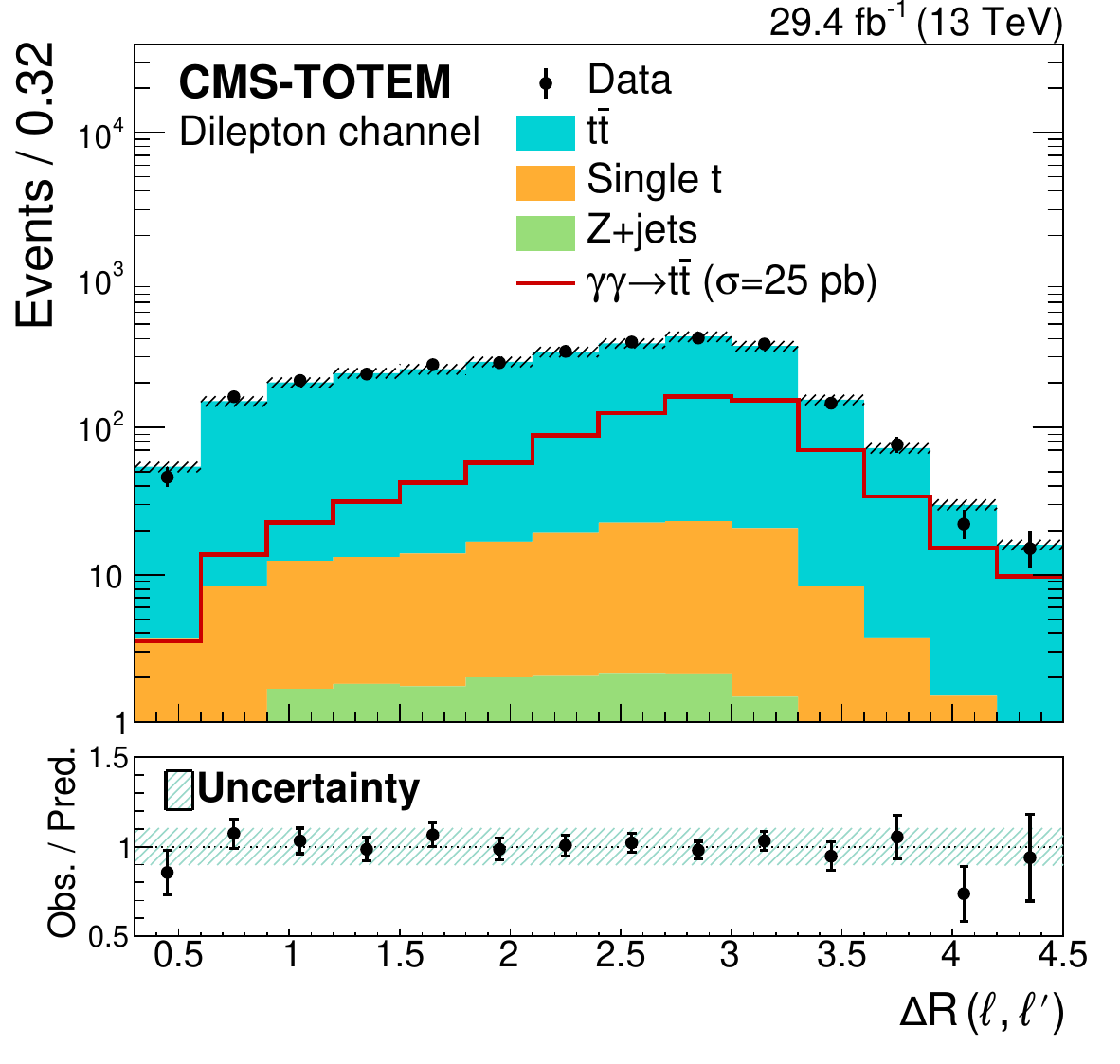}
  \includegraphics[width=0.32\textwidth]{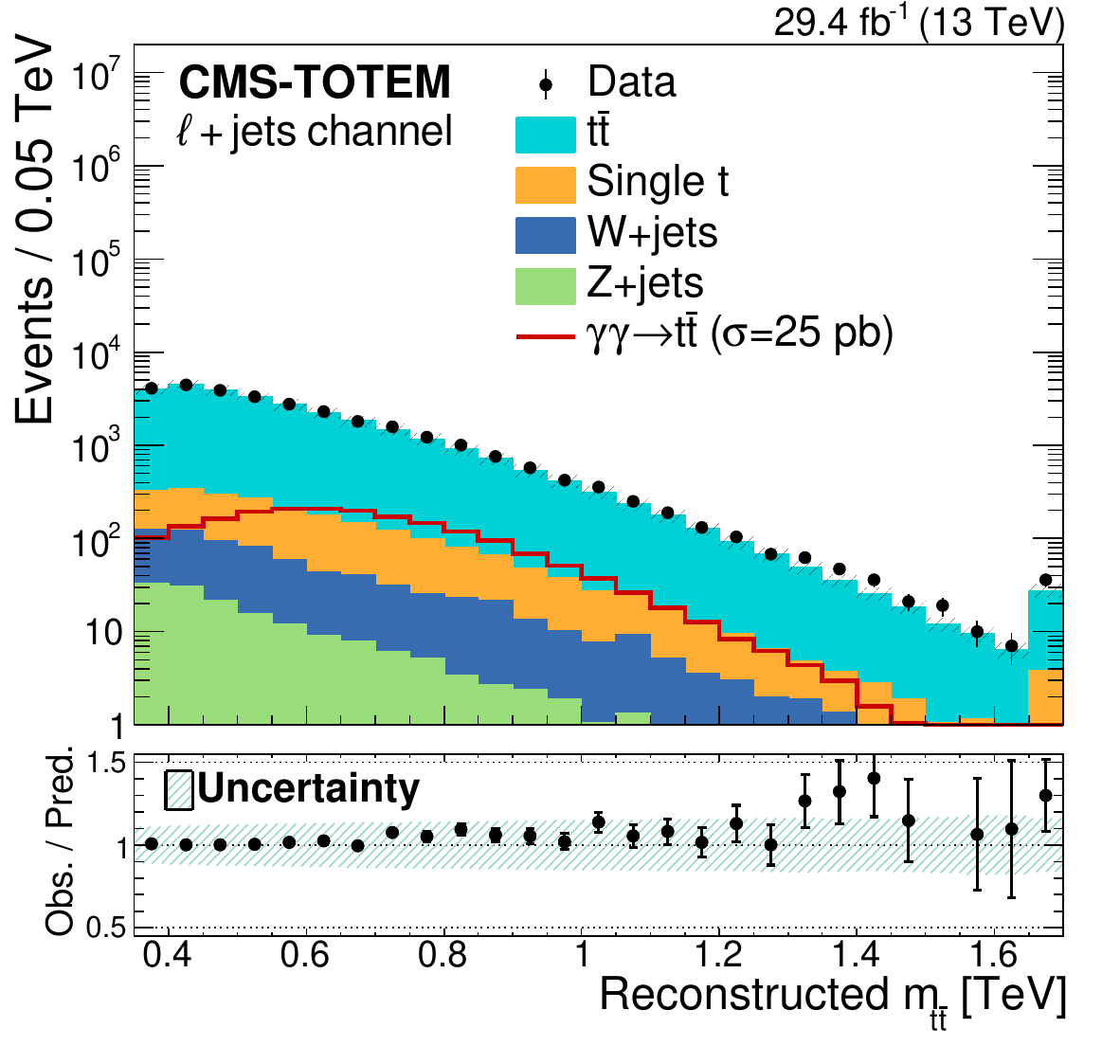}
  \includegraphics[width=0.32\textwidth]{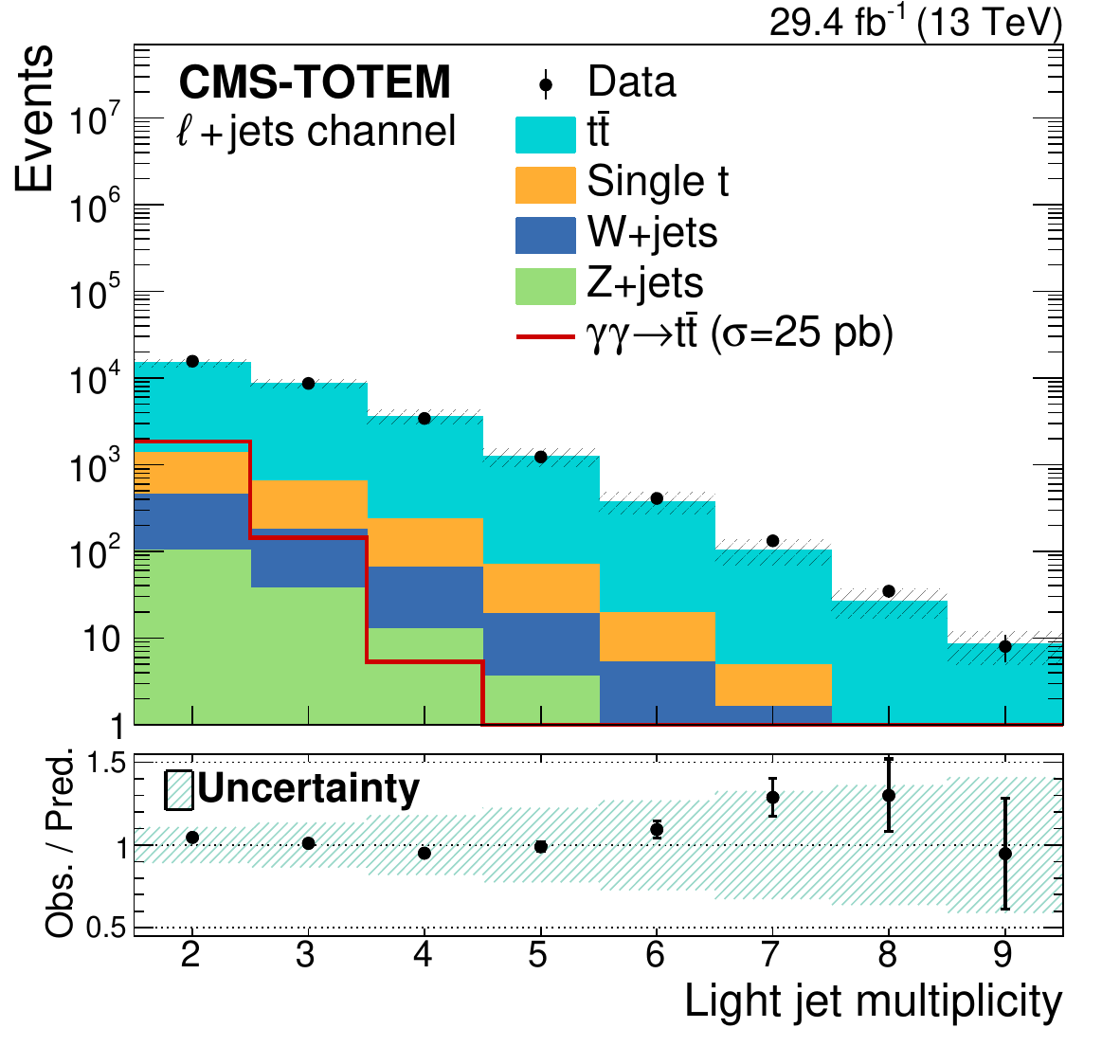}
  \includegraphics[width=0.32\textwidth]{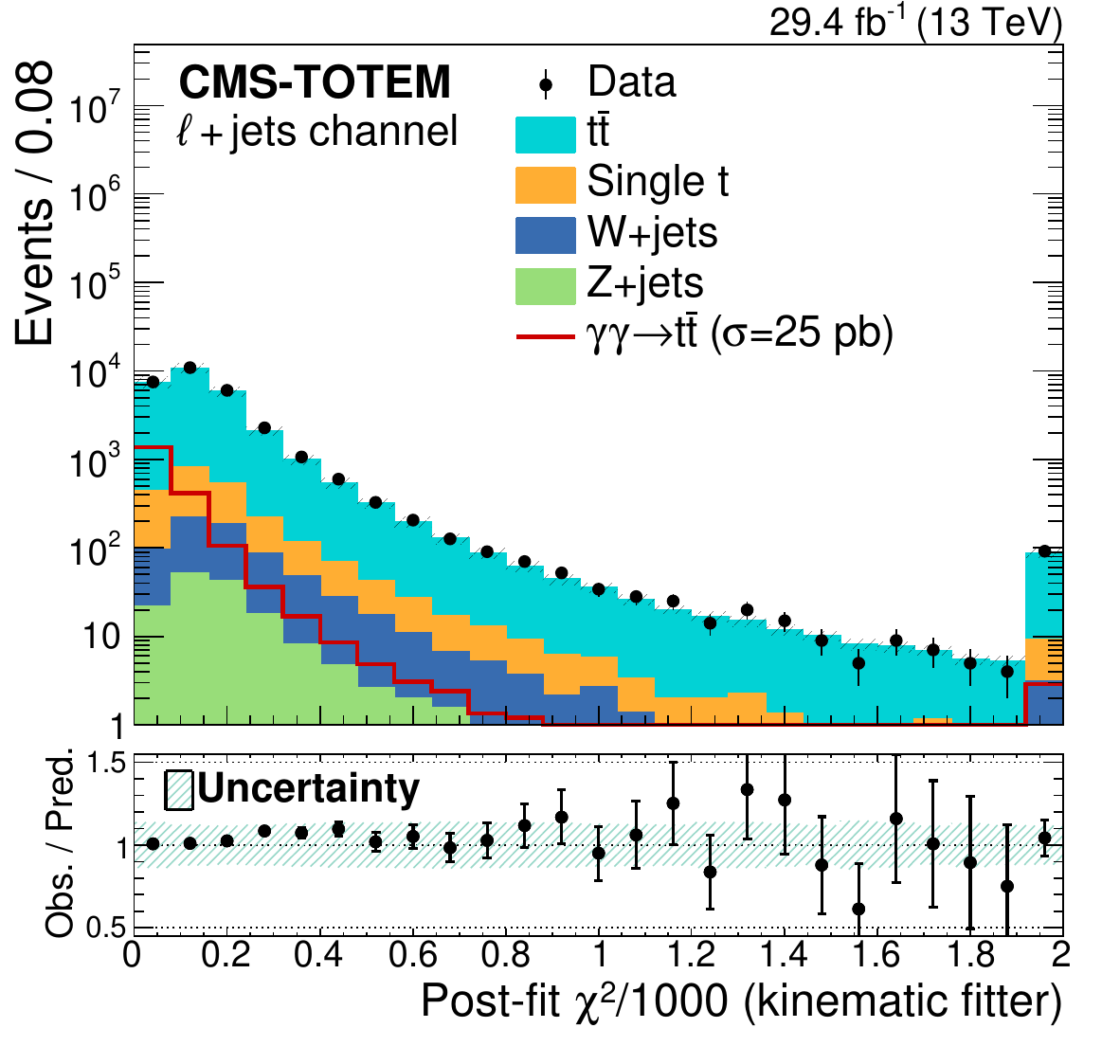}
  \caption{Distribution of a selection of the kinematic variables of interest for
    the dilepton (upper) and \ljets (lower) analysis.
    The solid histograms show the expected background contributions,
    while the red open histograms show the expected signal shapes, normalised
    to a cross section of 25\unit{pb}, approximately $10^5$ larger than the
    SM cross section prediction from Ref.~\cite{Shao:2022cly}; points
    with statistical error bars represent collision data.
    The lower panels show the data-to-prediction ratios; the hatched
    bands represent the uncertainty in the
    predictions. The leftmost and rightmost bin in each histogram includes
    accepted events outside the histogram range.}
  \label{fig:validationPlots}
\end{figure}

The distributions of some of the kinematic
variables of interest are shown in Fig.~\ref{fig:validationPlots} for
the two decay modes.

\section{Systematic uncertainties}
\label{sec:syst}

Several sources of systematic uncertainty affect the normalisation of
the signal and background yields, as well as the shape of the BDT
output used as the final discriminant. For each of them, the impact on
the final result is assessed by varying appropriately the parameters
involved, and repeating the analysis. When the variations imply a
change in the BDT shape, a smoothing procedure (using the `353QH'
algorithm described in Ref.~\cite{Friedman:1974vj}) is applied to the
associated template used in the fitting procedure described in
Section~\ref{sec:results}. Modified BDT shapes are compared to the
nominal one using a Kolmogorov--Smirnov-inspired test:
if the test result (calculated as described in Section~6.2.2 of
Ref.~\cite{Brun:2296378}) is larger than 0.95 for both the upwards and
downwards variation, the corresponding systematic uncertainty is only
included as an overall normalisation effect; otherwise, the
shape uncertainty is included as a nuisance parameter and profiled in
the likelihood fit.

The sources of systematic uncertainty can be subdivided into
experimental and theoretical components.

\paragraph*{Experimental uncertainties}
The measured integrated luminosity that is used to normalise the MC
predictions has an associated systematic uncertainty of
2.3\%~\cite{CMS:2018elu, CMS:LUM-17-003}. Several uncertainties arise from the
reconstruction and identification of various objects. For leptons, \PQb
quark jets, and forward protons, efficiency correction scale factors
are varied within their uncertainties~\cite{CMS:2020uim, CMS:2018rym, Khachatryan:2016kdb, TOTEM:2022vox},
which affect both the shape and
normalisation of the final discriminant. The uncertainty in the jet
energy has an effect on the reconstruction of the kinematic variables
used to calculate the discriminants: the corresponding uncertainty is
evaluated by rescaling the \pt- and $\eta$-dependent scale factors of
the reconstructed jet energy~\cite{Khachatryan:2016kdb} and jet energy
resolution. The variation in four-momentum for each selected jet is
propagated to \ptvecmiss and the \PQb tagging scale factors.
Uncertainties in the efficiency corrections for the lepton trigger are
estimated as functions of the lepton \pt and $\eta$ from control
samples in data; for electrons (muons) they are within 3\% (below
1\%), except for $\pt < 35\GeV$, where they range up to 8 (3)\%.
In the pileup proton mixing procedure described in
Section~\ref{subsec:models}, the normalisation of the simulated data
samples is performed according to the pileup proton probability measured in
real data with no requirement on the \PQb quark jet multiplicity. A
possible bias of the proton tag probability arising from the different \PQb
quark jet selection is estimated by measuring the proton tag probability
again after requiring $N_{\text{\PQb jet}}\geq 1$: the difference in the
predicted tagged proton probability is taken as the corresponding systematic
uncertainty. For the signal sample, the simulation of forward protons
is tuned to reproduce the expected bias and resolution in $\xi$
reconstruction assuming perfect knowledge of the detector alignment
and LHC optics. The effect of uncertainties in this assumption is
estimated by shifting, in each event, the reconstructed $\xi$ values
according to the ``systematics'' contribution described in Ref.~\cite{TOTEM:2022vox}.

\paragraph*{Theoretical uncertainties}
The uncertainties related to the choice of the factorisation and
renormalisation scales at the matrix element level are estimated by
varying the scales independently by factors 2 and 0.5
\cite{Cacciari:2003fi}. For PDF modeling, two effects are considered:
a variation of the strong coupling constant \alpS, and the
root-mean-square of the variations from a collection
of PDF error eigenvectors sets, as described in the PDF4LHC
Collaboration recommendations~\cite{Butterworth:2015oua}. The
uncertainty associated with parton shower emission in initial and final
state is evaluated by varying the renormalisation scale
for QCD emissions by factors of 2 and 0.5.
For the signal sample, only the final state radiation uncertainty is considered,
and is taken to be fully correlated with that of the background processes.
The normalisation of the inclusive \ttbar background (incorporating any additional
contribution from events with nonprompt leptons) is free to vary around its nominal
values for the \ljets and the dilepton channels
separately, while single top quark and other backgrounds normalisation
uncertainties are taken to be 5\%~\cite{Kidonakis:2021vob} and
30\%~\cite{CMS:2016eha, CMS:2020mxy, CMS:2021pcj}, respectively. Finally, the
effect of the finite size of the simulated samples used for the
analysis is taken into account with the Beeston--Barlow
method~\cite{Barlow:1993dm}.

\section{Results}
\label{sec:results}

A profile maximum-likelihood fit is performed to the distributions of
BDT discriminants for the two decay modes.
While the sensitivity with the current data does not allow to
obtain evidence for central exclusive \ttbar production, an upper limit
for its cross section can be derived.
The limits are computed
based on an asymptotic approximation of the distributions of the test
statistics, which in turn is based on the profile likelihood ratio,
under given hypotheses for the signal and the
background~\cite{Junk:1999kv, Read:2002hq, Cowan:2010js}. The sources of systematic
uncertainty described in Section~\ref{sec:syst} are included in the
fit as nuisance parameters.

The impact of a given systematic uncertainty on the upper limit is
defined as the relative difference between the nominal limit and the limit extracted by
including all other systematic uncertainties but excluding the
uncertainty in question. For the final result, uncertainties whose
impact on the upper limit is less than 0.1\% are not included.

In the dilepton analysis, a simultaneous fit is performed to each of the final-state lepton
combinations \ee, \emu, and \mumu, integrating over era and \alphaX. For the \ljets analysis,
the simultaneous fit is performed on each of the 20 samples defined by $(\text{era}, \alphaX)$,
combining the two lepton flavours.
These choices are the result of an optimisation based on a compromise
between the expected sensitivity and the statistical uncertainty.
The BDT distributions are binned in 14 and 22 intervals for the dilepton and the \ljets analysis,
respectively.

\begin{figure}[!htp]
  \centering
  \includegraphics[width=.46\textwidth]{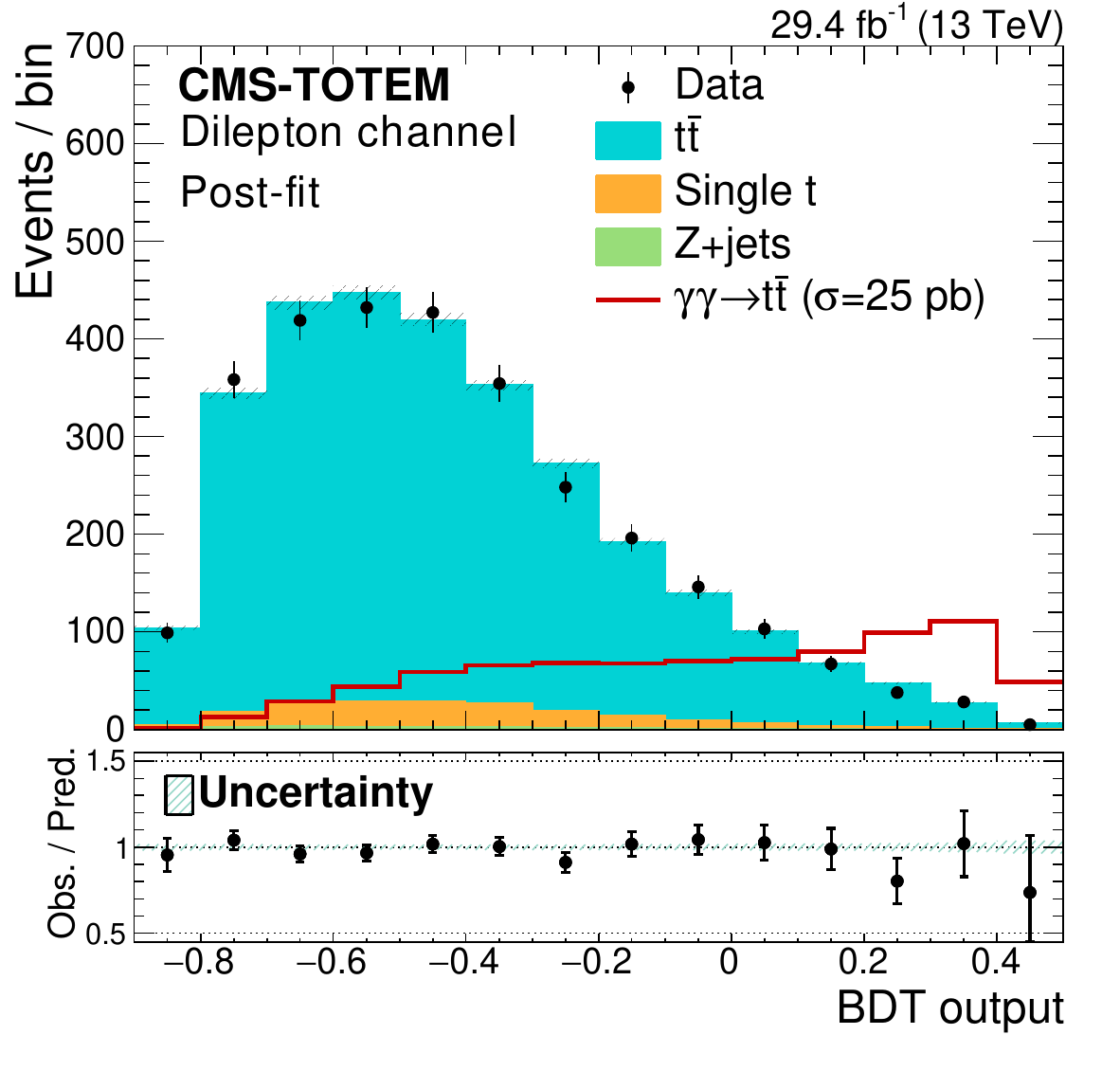}
  \includegraphics[width=.46\textwidth]{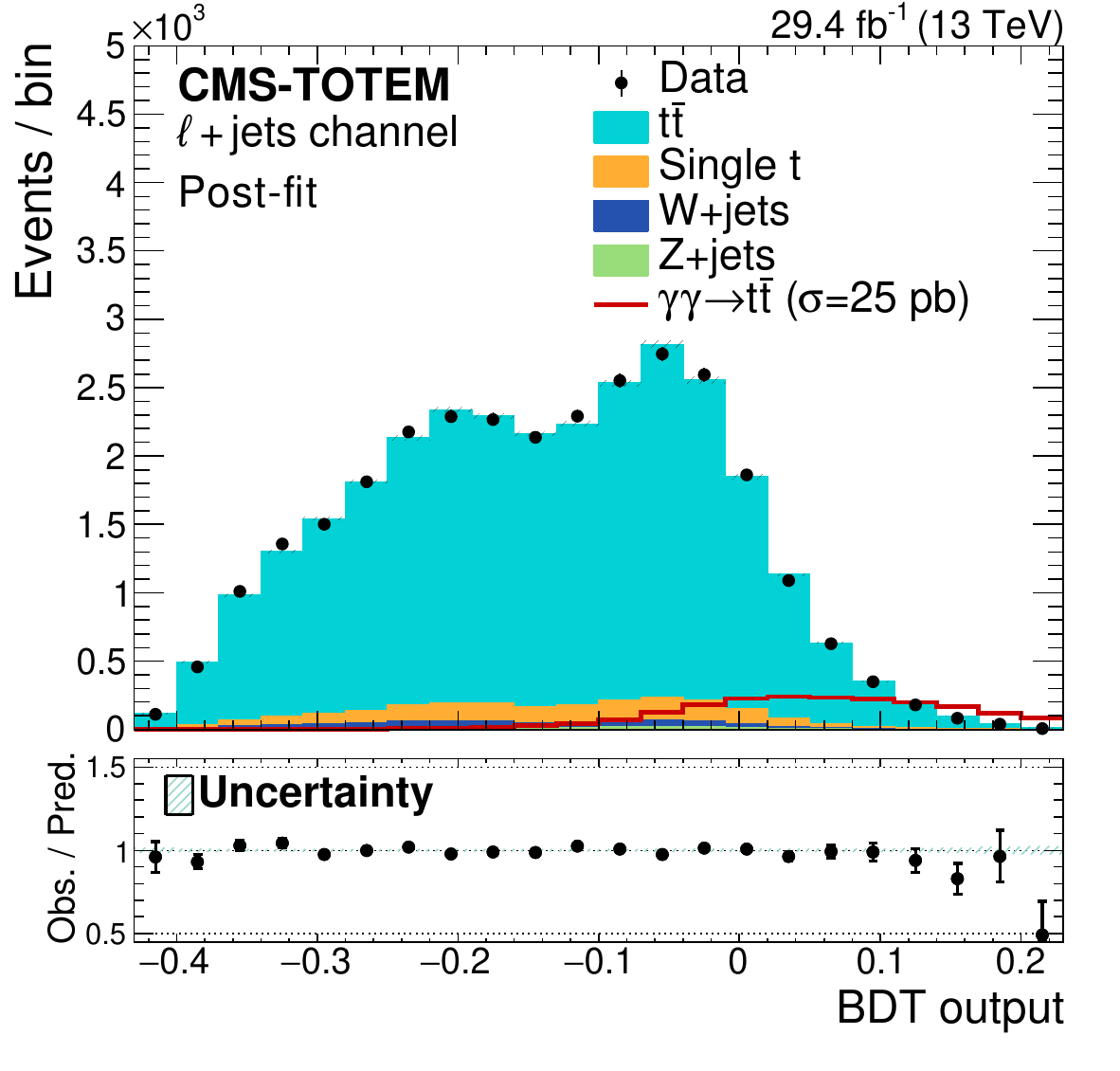}
  \caption{
    Distribution of the BDT output in the signal region for simulated events after the fit, and for data. Left: dilepton
    channel; right: \ljets channel. The different ranges of the two BDT output distributions
    depend on the specific details of the algorithms chosen in the two cases.
    The solid histograms show the expected background contributions,
    while the red open histograms show the expected signal shapes, normalised
    to a cross section of 25\unit{pb}, approximately $10^5$ larger than the
    SM cross section prediction from Ref.~\cite{Shao:2022cly}; points
    with statistical error bars represent collision data.
    For both reconstruction modes, all signal regions are combined.
    The lower panels show the data-to-prediction ratios; the hatched
    bands represent the relative uncertainty in the predictions.}
  \label{fig:BDT_final}
\end{figure}

The expected and observed distributions of the BDT
variable for the dilepton and \ljets decay modes are shown
in Fig.~\ref{fig:BDT_final}, where all signal regions are combined.
The values of the nuisance parameters returned by the fit are consistent
with their inputs; in particular, the normalisation factors for the \ttbar contribution to the
background are $0.96 \pm 0.04$ and $1.02 \pm 0.03$ for the dilepton and the \ljets channel,
respectively. The goodness-of-fit has been checked with toy-MC studies.

In the dilepton decay mode, the fit yields an observed (expected) 95\%
confidence level upper limit on exclusive central production of \ttbar
pairs of 1.71 (2.02)\unit{pb}; in the \ljets
mode, an upper limit of 0.78 (1.54)\unit{pb} is obtained.
The two modes are then considered jointly in a combined fit, where each
source of systematic uncertainty is treated as fully correlated between
the two channels. The observed (expected) limit resulting from the
combined fit is 0.59 (1.14)\unit{pb}.

\begin{figure}[!htp]
\centering
\includegraphics[width=.5\textwidth]{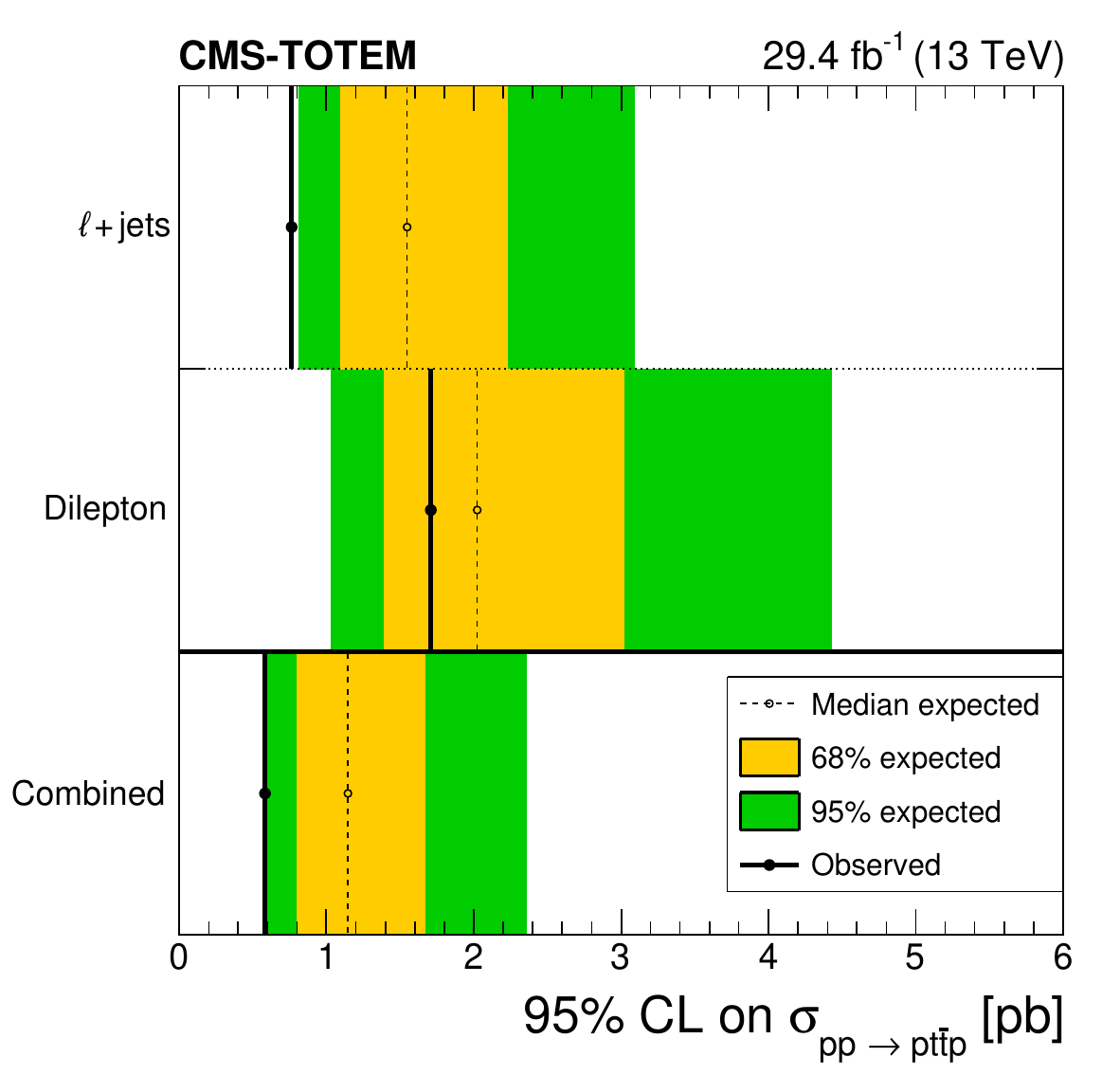}
\caption{Expected and observed 95\% confidence level (\CL) upper limits for the cross section of
  \pptopttbarp, for the dilepton and \ljets
  channels separately and combined. The green and yellow
  bands show the 68 and 95\% intervals, respectively, for the expected upper limit.}
\label{fig:limits}
\end{figure}

The results of the fit are shown in Fig.~\ref{fig:limits}, for the
separate decay channels, as well as for the combination.
The value of the extracted limit depends mostly on the statistical
precision; the increase due to inclusion of the systematic uncertainties is about 10\%.
The most important contributions from systematic uncertainties are those
related to background normalisation, final-state radiation modelling, jet
energy corrections and resolution, as well as proton reconstruction with CT-PPS.
These results are tabulated in the HEPData record for this analysis~\cite{hepdata}.

\section{Summary}
\label{sec:summary}

A search is reported for the central exclusive production of
top quark-antiquark pairs in proton-proton interactions, \pptopttbarp,
for the first time using tagged intact
protons, reconstructed by the CMS-TOTEM precision proton spectrometer.
The \ttbar pairs are reconstructed by the CMS detector either in the
dilepton or the \leptonjets decay modes. The search is conducted both separately
for the two modes, and in a combined fit. With a data sample of
proton-proton collisions at a centre-of-mass energy of 13\TeV
corresponding to an integrated luminosity of 29.4\fbinv, results
consistent with predictions from the standard model are obtained.
An upper limit of 0.59\unit{pb} at 95\% confidence level (compared to an expected limit of 1.14\unit{pb})
is set on the central exclusive production of \ttbar pairs, with
fractional momentum loss of the intact protons in the range $0.02 <
\xi < 0.20$.

\begin{acknowledgments}
We congratulate our colleagues in the CERN accelerator departments for the excellent performance of the LHC and thank the technical and administrative staffs at CERN and at other CMS and TOTEM institutes for their contributions to the success of the common CMS-TOTEM effort. In addition, we gratefully acknowledge the computing centres and personnel of the Worldwide LHC Computing Grid and other centres for delivering so effectively the computing infrastructure essential to our analyses. Finally, we acknowledge the enduring support for the construction and operation of the LHC, the CMS and TOTEM  detectors, and the supporting computing infrastructure provided by the following funding agencies: SC (Armenia), BMBWF and FWF (Austria); FNRS and FWO (Belgium); CNPq, CAPES, FAPERJ, FAPERGS, and FAPESP (Brazil); MES and BNSF (Bulgaria); CERN; CAS, MoST, and NSFC (China); MINCIENCIAS (Colombia); MSES and CSF (Croatia); RIF (Cyprus); SENESCYT (Ecuador); MoER, ERC PUT and ERDF (Estonia); Academy of Finland, Magnus Ehrnrooth Foundation, MEC,  HIP, and Waldemar von Frenckell Foundation (Finland); CEA and CNRS/IN2P3 (France); SRNSF (Georgia); BMBF, DFG, and HGF (Germany); GSRI (Greece); NKFIH (Hungary); DAE and DST (India); IPM (Iran); SFI (Ireland); INFN (Italy); MSIP and NRF (Republic of Korea); MES (Latvia); LAS (Lithuania); MOE and UM (Malaysia); BUAP, CINVESTAV, CONACYT, LNS, SEP, and UASLP-FAI (Mexico); MOS (Montenegro); MBIE (New Zealand); PAEC (Pakistan); MES and NSC (Poland); FCT (Portugal); MESTD (Serbia); MCIN/AEI and PCTI (Spain); MOSTR (Sri Lanka); Swiss Funding Agencies (Switzerland); MST (Taipei); MHESI and NSTDA (Thailand); TUBITAK and TENMAK (Turkey); NASU (Ukraine); STFC (United Kingdom); DOE and NSF (USA).

\hyphenation{Rachada-pisek} Individuals have received support from the Marie-Curie programme and the European Research Council and Horizon 2020 Grant, contract Nos.\ 675440, 724704, 752730, 758316, 765710, 824093, and COST Action CA16108 (European Union); the Leventis Foundation; the Alfred P.\ Sloan Foundation; the Alexander von Humboldt Foundation; the Science Committee, project no. 22rl-037 (Armenia); the Belgian Federal Science Policy Office; the Fonds pour la Formation \`a la Recherche dans l'Industrie et dans l'Agriculture (FRIA-Belgium); the Agentschap voor Innovatie door Wetenschap en Technologie (IWT-Belgium); the F.R.S.-FNRS and FWO (Belgium) under the ``Excellence of Science -- EOS" -- be.h project n.\ 30820817; the Beijing Municipal Science \& Technology Commission, No. Z191100007219010 and Fundamental Research Funds for the Central Universities (China); the Ministry of Education, Youth and Sports (MEYS) of the Czech Republic; Svenska Kulturfonden (Finland); the Shota Rustaveli National Science Foundation, grant FR-22-985 (Georgia); the Deutsche Forschungsgemeinschaft (DFG), under Germany's Excellence Strategy -- EXC 2121 ``Quantum Universe" -- 390833306, and under project number 400140256 - GRK2497; the Hellenic Foundation for Research and Innovation (HFRI), Project Number 2288 (Greece); the Hungarian Academy of Sciences, the New National Excellence Program - \'UNKP, the NKFIH research grants K 124845, K 124850, K 128713, K 128786, K 129058, K 131991, K 133046, K 138136, K 143460, K 143477, 2020-2.2.1-ED-2021-00181, and TKP2021-NKTA-64 (Hungary); the Council of Science and Industrial Research, India; the Latvian Council of Science; the Ministry of Education and Science, project no. 2022/WK/14, and the National Science Center, contracts Opus 2021/41/B/ST2/01369 and 2021/43/B/ST2/01552 (Poland); the Funda\c{c}\~ao para a Ci\^encia e a Tecnologia, grant CEECIND/01334/2018 (Portugal); the National Priorities Research Program by Qatar National Research Fund; MCIN/AEI/10.13039/501100011033, ERDF ``a way of making Europe", and the Programa Estatal de Fomento de la Investigaci{\'o}n Cient{\'i}fica y T{\'e}cnica de Excelencia Mar\'{\i}a de Maeztu, grant MDM-2017-0765 and Programa Severo Ochoa del Principado de Asturias (Spain); the Chulalongkorn Academic into Its 2nd Century Project Advancement Project, and the National Science, Research and Innovation Fund via the Program Management Unit for Human Resources \& Institutional Development, Research and Innovation, grant B05F650021 (Thailand); the Kavli Foundation; the Nvidia Corporation; the SuperMicro Corporation; the Welch Foundation, contract C-1845; and the Weston Havens Foundation (USA).
\end{acknowledgments}

\bibliography{auto_generated}
\cleardoublepage \appendix\section{The CMS Collaboration \label{app:collab}}\begin{sloppypar}\hyphenpenalty=5000\widowpenalty=500\clubpenalty=5000
\cmsinstitute{Yerevan Physics Institute, Yerevan, Armenia}
{\tolerance=6000
A.~Tumasyan\cmsorcid{0009-0000-0684-6742}
\par}
\cmsinstitute{Institut f\"{u}r Hochenergiephysik, Vienna, Austria}
{\tolerance=6000
W.~Adam\cmsorcid{0000-0001-9099-4341}, J.W.~Andrejkovic, T.~Bergauer\cmsorcid{0000-0002-5786-0293}, S.~Chatterjee\cmsorcid{0000-0003-2660-0349}, K.~Damanakis\cmsorcid{0000-0001-5389-2872}, M.~Dragicevic\cmsorcid{0000-0003-1967-6783}, A.~Escalante~Del~Valle\cmsorcid{0000-0002-9702-6359}, P.S.~Hussain\cmsorcid{0000-0002-4825-5278}, M.~Jeitler\cmsAuthorMark{1}\cmsorcid{0000-0002-5141-9560}, N.~Krammer\cmsorcid{0000-0002-0548-0985}, L.~Lechner\cmsorcid{0000-0002-3065-1141}, D.~Liko\cmsorcid{0000-0002-3380-473X}, I.~Mikulec\cmsorcid{0000-0003-0385-2746}, P.~Paulitsch, F.M.~Pitters, J.~Schieck\cmsAuthorMark{1}\cmsorcid{0000-0002-1058-8093}, R.~Sch\"{o}fbeck\cmsorcid{0000-0002-2332-8784}, D.~Schwarz\cmsorcid{0000-0002-3821-7331}, S.~Templ\cmsorcid{0000-0003-3137-5692}, W.~Waltenberger\cmsorcid{0000-0002-6215-7228}, C.-E.~Wulz\cmsAuthorMark{1}\cmsorcid{0000-0001-9226-5812}
% ...
\par}
\vskip\cmsinstskip
\dag:~Deceased\\
$^{1}$Also at TU Wien, Vienna, Austria\\
% ...

\section{The TOTEM Collaboration\label{app:totem}}
\newcommand{\AddAuthor}[2]{#1$^{#2}$,\ }
\newcommand\AddAuthorLast[2]{#1$^{#2}$}
\noindent
\AddAuthor{G.~Antchev\cmsorcid{0000-0003-3210-5037}}{a}
\AddAuthor{P.~Aspell}{8}
\AddAuthor{I.~Atanassov\cmsorcid{0000-0002-5728-9103}}{a}
\AddAuthor{V.~Avati}{7,8}
\AddAuthor{J.~Baechler}{8}
\AddAuthor{C.~Baldenegro~Barrera\cmsorcid{0000-0002-6033-8885}}{11}
\AddAuthor{V.~Berardi\cmsorcid{0000-0002-8387-4568}}{4a,4b}
\AddAuthor{M.~Berretti\cmsorcid{0000-0003-4122-8282}}{2a}
\AddAuthor{V.~Borshch\cmsorcid{0000-0002-5479-1982}}{12}
\AddAuthor{E.~Bossini\cmsorcid{0000-0002-2303-2588}}{6a}
\AddAuthor{U.~Bottigli\cmsorcid{0000-0002-0666-3433}}{6b}
\AddAuthor{M.~Bozzo\cmsorcid{0000-0002-1715-0457}}{5a,5b}
\AddAuthor{H.~Burkhardt}{8}
\AddAuthor{F.S.~Cafagna\cmsorcid{0000-0002-7450-4784}}{4a}
\AddAuthor{M.G.~Catanesi\cmsorcid{0000-0002-2987-7688}}{4a}
\AddAuthor{M.~Deile\cmsorcid{0000-0001-5085-7270}}{8}
\AddAuthor{F.~De~Leonardis}{4a,4c}
\AddAuthor{M.~Doubek\cmsorcid{0000-0002-0026-9558}}{1c}
\AddAuthor{D.~Druzhkin}{3a,8}
\AddAuthor{K.~Eggert}{10}
\AddAuthor{V.~Eremin}{c}
\AddAuthor{A.~Fiergolski}{8}
\AddAuthor{F.~Garcia\cmsorcid{0000-0002-4023-7964}}{2a}
\AddAuthor{V.~Georgiev}{1a}
\AddAuthor{S.~Giani}{8}
\AddAuthor{L.~Grzanka\cmsorcid{0000-0002-3599-854X}}{7}
\AddAuthor{J.~Hammerbauer}{1a}
\AddAuthor{T.~Isidori\cmsorcid{0000-0002-7934-4038}}{11}
\AddAuthor{V.~Ivanchenko\cmsorcid{0000-0002-1844-5433}}{12}
\AddAuthor{M.~Janda\cmsorcid{0000-0001-5736-6183}}{1c}
\AddAuthor{A.~Karev}{8}
\AddAuthor{J.~Ka\v{s}par\cmsorcid{0000-0001-5639-2267}}{1b,8}
\AddAuthor{B.~Kaynak\cmsorcid{0000-0003-3857-2496}}{9}
\AddAuthor{J.~Kopal\cmsorcid{0000-0001-9751-7409}}{8}
\AddAuthor{V.~Kundr\'{a}t\cmsorcid{0000-0003-2868-5550}}{1b}
\AddAuthor{S.~Lami\cmsorcid{0000-0001-9492-0147}}{6a}
\AddAuthor{R.~Linhart}{1a}
\AddAuthor{C.~Lindsey}{11}
\AddAuthor{M.V.~Lokaj\'{i}\v{c}ek\cmsorcid{0000-0002-2052-1220}$^{\textrm{\dag,}}$}{1b}
\AddAuthor{L.~Losurdo\cmsorcid{0000-0002-4964-7951}}{6b}
\AddAuthor{F.~Lucas~Rodr\'{i}guez}{8}
\AddAuthor{M.~Macr\'{i}}{\dag,5a}
\AddAuthor{M.~Malawski\cmsorcid{0000-0001-6005-0243}}{7}
\AddAuthor{N.~Minafra\cmsorcid{0000-0003-4002-1888}}{11}
\AddAuthor{S.~Minutoli}{5a}
\AddAuthor{K.~Misan\cmsorcid{0009-0007-2416-042X}}{7}
\AddAuthor{T.~Naaranoja\cmsorcid{0000-0001-5797-7929}}{2a,2b}
\AddAuthor{F.~Nemes\cmsorcid{0000-0002-1451-6484}}{3a,3b,8}
\AddAuthor{H.~Niewiadomski}{10}
\AddAuthor{E.~Oliveri\cmsorcid{0000-0002-0832-6975}}{8}
\AddAuthor{F.~Oljemark\cmsorcid{0000-0003-0121-2761}}{2a,2b}
\AddAuthor{M.~Oriunno}{b}
\AddAuthor{K.~\"{O}sterberg\cmsorcid{0000-0003-4807-0414}}{2a,2b}
\AddAuthor{S.~Ozkorucuklu\cmsorcid{0000-0001-5153-9266}}{9}
\AddAuthor{P.~Palazzi\cmsorcid{0000-0002-4861-391X}}{8}
\AddAuthor{V.~Passaro\cmsorcid{0000-0003-0802-4464}}{4a,4c}
\AddAuthor{Z.~Peroutka}{1a}
\AddAuthor{O.~Potok\cmsorcid{0009-0005-1141-6401}}{9}
\AddAuthor{J.~Proch\'{a}zka\cmsorcid{0000-0002-2774-2245}}{1b}
\AddAuthor{M.~Quinto\cmsorcid{0000-0002-6363-6132}}{4a,4b}
\AddAuthor{E.~Radermacher}{8}
\AddAuthor{E.~Radicioni\cmsorcid{0000-0002-2231-1067}}{4a}
\AddAuthor{F.~Ravotti\cmsorcid{0000-0002-5709-6934}}{8}
\AddAuthor{C.~Royon\cmsorcid{0000-0002-7672-9709}}{11}
\AddAuthor{G.~Ruggiero}{8}
\AddAuthor{H.~Saarikko}{2a,2b}
\AddAuthor{V.D.~Samoylenko}{c}
\AddAuthor{A.~Scribano\cmsorcid{0000-0002-4338-6332}}{6a,11}
\AddAuthor{J.~\v{S}irok\'{y}}{1a}
\AddAuthor{J.~Smajek}{8}
\AddAuthor{W.~Snoeys\cmsorcid{0000-0003-3541-9066}}{8}
\AddAuthor{R.~Stefanovitch}{3a,8}
\AddAuthor{C.~Taylor\cmsorcid{0000-0001-6816-8051}}{10}
\AddAuthor{E.~Tcherniaev\cmsorcid{0000-0002-3685-0635}}{12}
\AddAuthor{N.~Turini\cmsorcid{0000-0002-9395-5230}}{6b}
\AddAuthor{O.~Urban}{1a}
\AddAuthor{V.~Vacek\cmsorcid{0000-0001-9584-0392}}{1c}
\AddAuthor{O.~Vavroch}{1a}
\AddAuthor{J.~Welti}{2a,2b}
\AddAuthor{J.~Williams\cmsorcid{0000-0002-9810-7097}}{11}
\AddAuthorLast{J.~Zich}{1a}
\vskip 4pt plus 4pt
\let\thefootnote\relax
\newcommand{\AddInstitute}[2]{${}^{#1}$#2\\}
\newcommand{\AddExternalInstitute}[2]{\footnote{${}^{#1}$ #2}}
\noindent
\dag Deceased\\
\AddInstitute{1a}{University of West Bohemia, Pilsen, Czech Republic}
\AddInstitute{1b}{Institute of Physics of the Academy of Sciences of the Czech Republic, Prague, Czech Republic}
\AddInstitute{1c}{Czech Technical University, Prague, Czech Republic}
\AddInstitute{2a}{Helsinki Institute of Physics, University of Helsinki, Helsinki, Finland}
\AddInstitute{2b}{Department of Physics, University of Helsinki, Helsinki, Finland}
\AddInstitute{3a}{Wigner Research Centre for Physics, RMKI, Budapest, Hungary}
\AddInstitute{3b}{MATE Institute of Technology KRC, Gy\"{o}ngy\"{o}s, Hungary}
\AddInstitute{4a}{INFN Sezione di Bari, Bari, Italy}
\AddInstitute{4b}{Dipartimento Interateneo di Fisica di Bari, University of Bari, Bari, Italy}
\AddInstitute{4c}{Dipartimento di Ingegneria Elettrica e dell'Informazione --- Politecnico di Bari, Bari, Italy}
\AddInstitute{5a}{INFN Sezione di Genova, Genova, Italy}
\AddInstitute{5b}{Universit\`{a} degli Studi di Genova, Genova, Italy}
\AddInstitute{6a}{INFN Sezione di Pisa, Pisa, Italy}
\AddInstitute{6b}{Universit\`{a} degli Studi di Siena and Gruppo Collegato INFN di Siena, Siena, Italy}
\AddInstitute{7}{Akademia G\'{o}rniczo-Hutnicza (AGH) University of Krak\'ow, Faculty of Computer Science, Krak\'ow, Poland}
\AddInstitute{8}{CERN, Geneva, Switzerland}
\AddInstitute{9}{Istanbul University, Istanbul, Turkey}
\AddInstitute{10}{Case Western Reserve University, Department of Physics, Cleveland, Ohio, USA}
\AddInstitute{11}{The University of Kansas, Lawrence, Kansas, USA}
\AddInstitute{12}{Authors affiliated with an institute or an international laboratory covered by a cooperation agreement with CERN}
\AddExternalInstitute{a}{INRNE-BAS, Institute for Nuclear Research and Nuclear Energy, Bulgarian Academy of Sciences, Sofia, Bulgaria}
\AddExternalInstitute{b}{SLAC, Stanford University, California, USA}
\AddExternalInstitute{c}{Authors affiliated with an institute or an international laboratory covered by a cooperation agreement with CERN}\end{sloppypar}
\end{document}